\documentclass[a4paper,aps,onecolumn,nofootinbib,10pt]{revtex4}

\usepackage{amsmath}
\usepackage{slashed}
\usepackage[english,]{babel}
\usepackage {float}
\usepackage{braket}
\usepackage{enumitem}
\usepackage{makeidx,amssymb,xspace}
\usepackage{graphicx}
\usepackage{epsfig,amsfonts,amssymb,amsmath}
\usepackage{cancel}
\usepackage{pgfplots} 
\usepackage{mathrsfs}
\usepackage{extsizes}
\usepackage[a4paper,top=4cm,bottom=4cm,left=2cm,right=2cm]{geometry}

\RequirePackage[colorlinks,hyperindex]{hyperref}
\hypersetup{colorlinks=true,breaklinks=true,urlcolor=blue,linkcolor=red}

\makeindex

\newcommand{\be}{\begin{equation}}\newcommand{\ee}{\end{equation}}
\newcommand{\bea}{\begin{eqnarray}}\newcommand{\eea}{\end{eqnarray}}
\newcommand{\brr}{\begin{array}}\newcommand{\err}{\end{array}}
\newcommand{\bit}{\begin{itemize}}\newcommand{\eit}{\end{itemize}}
\newcommand{\ben}{\begin{enumerate}}\newcommand{\een}{\end{enumerate}}

\def\non{\nonumber}

\def\non{\nonumber}

\def\1{{_{1}}}\def\2{{_{2}}}

\begin{document}
%%%%%%%%%%%%%%%%%%%%%%%%%%%%%%%%%%%%%%%%%%%%%%%%%%%%%%%%%%%%%%%%%%%%%%%%%%%%%%%%%%%%%%%%%%%%%%%%%%%
\title{A covariant approach to the Dirac field in LRS space-times}
\author{Stefano Vignolo,$^{c}$\!\! $^{\hbar}$\!\! $^{G}$\footnote{stefano.vignolo@unige.it}Giuseppe De Maria,$^{c}$\!\! $^{\hbar}$\footnote{giuseppe.demaria@edu.unige.it}Luca Fabbri,$^{c}$\!\! $^{\hbar}$\!\! $^{G}$\footnote{luca.fabbri@unige.it}Sante Carloni,$^{c}$\!\! $^{\hbar}$\!\! $^{G}$\!\! $^{\nabla}$\footnote{sante.carloni@unige.it}}
\affiliation{$^{c}$DIME, Universit\`{a} di Genova, Via all'Opera Pia 15, 16145 Genova, ITALY\\
$^{\hbar}$INFN, Sezione di Genova, Via Dodecaneso 33, 16146 Genova, ITALY\\
$^{G}$GNFM, Istituto Nazionale di Alta Matematica, P.le Aldo Moro 5, 00185 Roma, ITALY\\
$^{\nabla}$Institute of Theoretical Physics, Faculty of Mathematics and Physics,\\
Charles University, Prague, V Hole\v{s}ovi\v{c}k\'{a}ch 2, 180 00 Prague 8, CZECH REPUBLIC}
\date{\today}
%%%%%%%%%%%%%%%%%%%%%%%%%%%%%%%%%%%%%%%%%%%%%%%%%%%%%%%%%%%%%%%%%%%%%%%%%%%%%%%%%%%%%%%%%%%%%%%%%%%
\begin{abstract}
We use the polar decomposition to describe the Dirac field in terms of an effective spinorial fluid. After reformulating all covariant equations in ``spinorial'' signature $(+---)$, we develop a $(1+1+2)$ covariant approach for the Dirac field that does not require the use of tetrad fields or Clifford matrices. By identifying the velocity and spin fields as the generators of time-like and space-like congruences, we examine the compatibility of a self-gravitating Dirac field with Locally Rotationally Symmetric space-times of types I, II, and III. We provide illustrative examples to demonstrate the effectiveness of our construction.
\\
\\
{\bf keywords:} Covariant formalisms, polar decomposition, LRS space-times, self--gravitating Dirac field.
\end{abstract}
%%%%%%%%%%%%%%%%%%%%%%%%%%%%%%%%%%%%%%%%%%%%%%%%%%%%%%%%%%%%%%%%%%%%%%%%%%%%%%%%%%%%%%%%%%%%%%%%%%%
\maketitle
%%%%%%%%%%%%%%%%%%%%%%%%%%%%%%%%%%%%%%%%%%%%%%%%%%%%%%%%%%%%%%%%%%%%%%%%%%%%%%%%%%%%%%%%%%%%%%%%%%%
%%%%%%%%%%%%%%%%%%%%%%%%%%%%%%%%%%%%%%%%%%%%%%%%%%%%%%%%%%%%%%%%%%%%%%%%%%%%%%%%%%%%%%%%%%%%%%%%%%%
\section{Introduction} \label{Section1}
When dealing with relativistic gravitational systems, one often finds that an approach based on coordinates is not the most effective option.  The reason behind the shortcoming of the initial choice of coordinates is connected to the fact that (i) a coordinate system carries a number of hidden choices on the motion of the observers that use such coordinates, which might not be suitable to describe a given phenomenon and (ii) the inherent limitations of the coordinate system itself in terms of their regularity throughout the space-time manifold. A classical example of this situation is the Schwarzschild metric written in the classical Schwarzschild coordinates. These coordinates assume an observer that is static and very far from the source, which is not necessarily useful for describing physical processes that involve changes in the observer's position, such as the Oppenheimer-Snyder collapse (i). Moreover, they are singular on the horizon even if no true space-time singularity is present (ii).

These considerations help us to understand why the research community was led to investigate alternative ways to analyze space-times that rely only marginally on the choice of a coordinate system. One of the most famous examples is the Arnowitt-Deser-Misner (ADM) formalism \cite{ADM}, in which the metric is expressed in terms of a scalar (lapse), a vector (shift), and a three-dimensional metric, in turn, connected with the extrinsic curvature. This decomposition enables us to write the Einstein equations, or, more commonly, to construct the Hamiltonian that generates these equations in terms of these quantities. Another approach of this type is the Newman-Penrose (NP) formalism \cite{NP} in which the idea is to project all tensors of the theory onto a base made of four $4$-vectors, two of which are null. The twelve scalars obtained in this way, called spin coefficients, are then used to write the gravitational field equations as a system of scalar equations. The NP formalism can be seen as a covariant generalization of the well-known tetrad formalism, which, however, does not rely too heavily on a frame choice.

Much more recently, on the basis of Ehlers' work on relativistic hydrodynamics \cite{Ehlers:1961xww}, Ellis and coworkers introduced the so-called covariant approaches \cite{cargese,Ellisperf,Clarkson}.  These formalisms exploit as much as possible the symmetries that a given class of space-times might have by choosing up to two $4$-vectors and employing a suitable foliation based on them. Two versions of these formalisms have been employed so far. The first, called the $(1+3)$ covariant approach, uses only one time-like $4$-vector, and a second one, called the $(1+1+2)$ covariant approach, employs a time-like and a space-like $4$-vector. The covariant formulations have features similar to both the ADM and NP formalisms. Similarly to the first, they employ a foliation to characterize the space-time in terms of scalars, vectors, and tensors defined on a lower dimensional subspace of the space-time manifold, and the (decomposition of) the extrinsic curvature plays a crucial role. However, such decomposition is associated with the choice of some specific $4$-vectors much in the same way as the NP formalism, albeit these $4$-vectors are time-like or space-like and are usually only one or two. Another, and probably the most important, difference between the NP formalism (and in part the ADM one) and the covariant approach is that the choice of the $4$-vectors, as well as the quantities involved, has a very clear physical interpretation, thereby helping the understanding of complex space-times, useful in both astrophysics and cosmology.

Covariant formalisms have been employed successfully for several problems in cosmology and astrophysics in Einstein gravity, like the generalization of the Ehlers-Geren-Sachs theorem \cite{1995ApJ...443....1S}, the construction of a covariant gauge-invariant theory of perturbations for cosmology \cite{bed, eb, ehb, ebh}, black holes \cite{Clarkson}, and more recently relativistic stars \cite{Carloni:2017rpu,Carloni:2017bck,Naidu:2022igk,Naidu:2021nwh,Luz:2024yjm,Luz:2024lgi,Luz:2024xnd}.

Although covariant approaches are designed with the idea that the source of the gravitational field is one or more traditional fluid continua, the extension to treat classical fields has been attempted with success at both the exact and perturbative levels. For example, in the case of scalar fields, both cosmological perturbations \cite{Carloni:2006gy} and an extension of Derrick's theorem \cite{Carloni:2019cyo} have been proposed. The reason for this success lies in the ability to represent these fields as effective fluids, thanks to the structure of their energy-momentum tensor.

The situation is different for fermionic fields $\psi$. In this case, the presence of Clifford matrices and non-trivial derivatives of spinors makes the standard application of the covariant formalism hard to develop. Things may be simpler when spinors are in plane-waves, because in this case the covariant derivative $\nabla_{k}\psi=-iP_{k}\psi$ is proportional to the spinor itself, all quantities can be reduced to spinorial bi-linears straightforwardly, and computations become manageable. However, the assumption of planar waves for spinors, interpretable with the fact that fermionic particles are to be considered point-like, while impressively accurate to treat high-energy scattering, is of no use for other systems (for example, the hydrogen atom itself does not have solutions in the form of plane waves). For these general systems, the covariant derivative $\nabla_{k}\psi$ is not simply proportional to the spinor itself \cite{Fabbri:2024eec}, making the covariant splitting impossible. This impasse may be circumvented by writing spinors in the so-called polar form. 

In polar form, spinors are basically written as the product of a modulus times phases in such a way that manifest covariance is preserved \cite{jl1, jl2}. When this polar form is implemented also at the differential level, it becomes possible to perform the polar decomposition of the Dirac equation \cite{tr1, tr2}.

Writing spinors in polar form has several advantages. A first is that the $4$ complex, or $8$ real, components of the spinor are re-configured into a set of variables given by the density distribution, the velocity, the spin, and a chiral angle, which can be recognized as an enlarged set of hydrodynamic variables \cite{t2}. The ensuing Dirac equations in polar form can be seen as a type of field equations for a fluid with spin \cite{Fabbri:2023onb}. A second advantage is that in polar form, there remain no explicit tetrads, or gamma matrices, so that there is less dependence on peculiar ways of representing spinors, and hence a higher generality. A third and most important advantage is that the covariant derivative of the spinor can be expressed in the form $\nabla_{k}\psi=M_{k}\psi$ for some matrix $M_{k}$ \cite{Fabbri:2023dgv}: this means that in polar form the spinor field's covariant derivative is indeed proportional to the spinor itself, and the covariant splitting becomes again possible \cite{Fabbri:2025ffi}.

By exploiting the main features and the consequent advantages offered by the polar decomposition, we propose a preliminary attempt at a covariant formulation of the self-gravitating Dirac field, without resorting to the tetrad formalism. The starting point of the proposed construction is the chance of describing the Dirac field in hydrodynamic terms, involving only real tensorial quantities. The next step consists of observing that the velocity and spin (pseudo) vector fields of the Dirac field naturally generate the two time-like and space-like congruences that form the fundamental elements of the $(1+1+2)$ covariant splitting. Before directly applying the $(1+1+2)$ decomposition, two preliminary steps are necessary. The first is the formulation of the covariant equations using the signature $(+---)$, which is the one commonly adopted in the treatment of spinor fields. The second involves performing the $(1+1+2)$ decomposition of both the energy--momentum tensor of the Dirac field and the Dirac equation itself, after expressing them in polar form.

Once this has been done, the entire geometrical framework of covariant approaches is ready to be applied. In particular, here we focus on spinor fields in backreaction with Locally Rotationally Symmetric (LRS) space-times of types I, II, and III. We discuss the compatibility of the Dirac field with LRS geometries, both in the case where the spinor fluid is perfect and in the non-perfect case, under the assumption that the velocity and spin fields of the Dirac field are the generators of the time-like and space-like congruences of the $(1+1+2)$ covariant splitting. We also give some examples where the resulting covariant equations can be solved analytically.

The layout of the paper is as follows. Section \ref{Section2} presents the covariant equations of the $(1+3)$ and $(1+1+2)$ decompositions in the signature $(+---)$. Section \ref{Section3} discusses the consistency and integrability conditions for the so-obtained covariant equations. Section \ref{Section4} briefly reviews the main features of the polar formalism and implements the $(1+1+2)$ covariant decomposition of the energy--momentum tensor of the spinor field and the Dirac equations, all expressed in polar form. Section \ref{Section5} realizes the matching between polar formalism and covariant approach, providing a $(1+1+2)$ covariant formulation of the self-gravitating Dirac field in LRS space-times of type I, II, and III. Section \ref{Section6} illustrates some exact solutions.

Throughout the paper natural units ($c=8\pi G=1$) and metric signature $(+---)$ are used. Einstein's equations are written as
$$
G_{ab}=T_{ab}
$$ 
where $G_{ab}$ and $T_{ab}$ are the Einstein and the energy--momentum tensors. The Riemann tensor for $4$-dimensional space-time is expressed as
$$
R^{a}{}_{bcd} := \partial_{c}\Gamma_{db}{}^{a} - \partial_{d}\Gamma_{cb}{}^{a} + \Gamma_{cp}{}^{a}\Gamma_{db}{}^{p} - \Gamma_{dp}{}^{a}\Gamma_{cb}{}^{p}
$$
where $\Gamma_{ab}{}^{c}\partial_c:=\nabla_{\partial_a}\partial_b$, $\nabla$ denoting the (Levi--Civita) covariant derivative. The Ricci tensor is defined as $R_{ab}:=R^{c}{}_{acb}$. The symmetrization and antisymmetrization of expressions with two indexes are given by $W_{(ab)}:=\frac{1}{2}\left(W_{ab}+W_{ba}\right)$ and $W_{[ab]}:=\frac{1}{2}\left(W_{ab}-W_{ba}\right)$.

%%%%%%%%%%%%%%%%%%%%%%%%%%%%%%%%%%%%%%%%%%%%%%%%%%%%%%%%%%%%%%%%%%%%%%%%%%%%%%%%%%%%%%%%%%%%%%%%%
%%%%%%%%%%%%%%%%%%%%%%%%%%%%%%%%%%%%%%%%%%%%%%%%%%%%%%%%%%%%%%%%%%%%%%%%%%%%%%%%%%%%%%%%%%%%%%%%%
\section{Covariant formalism in signature $(+---)$} \label{Section2}
Since its first formulation, the covariant formalism has been developed by adopting the signature $(-+++)$. On the other hand, the signature $(+---)$ is commonly used when dealing with spinorial fields. Therefore, in order to implement a covariant approach to the Dirac field, we need the covariant equations concerning the $(1+3)$ and $(1+1+2)$ splittings in the signature $(+---)$. In this preliminary section, we present such equations, which, to the best of our knowledge, are not found in the literature. As expected, the covariant equations in signature $(+---)$ differ from those in signature $(-+++)$ by some signs. For brevity, we will omit the details of the explicit deduction of these equations. Besides, calculations are straightforward, although somewhat lengthy in some cases. Definitions and notations are borrowed from \cite{cargese,Ellisperf,Clarkson,Ellisnonperf}.

\subsection{$(1+3)$ covariant equations in signature $(+---)$} \label{Subsection2.1}
We denote by $u^{a}$ the unit $4$-vector of an assigned time-like congruence which represents the world lines of given observers. Here indices run from $0$ to $3$ ($a=0,\ldots,3$) and we have $u^{a}u_{a}=1$. Given the space-time metric $g_{ab}$, the projection operators
\bea\label{projectiom_operators}
U^{a}\hspace{0.1mm}_{b}:=u^{a}u_{b}\ \ \ \ {\rm and} \ \ \ \ h^a\hspace{0.1mm}_{b}:=g^a\hspace{0.1mm}_{b}- u^{a}u_{b} 
\eea 
allow us to decompose vectors (or, more generally, tensors) into components parallel and orthogonal to $u^{a}$, respectively. They satisfy the relations
\bea\label{rel_op_1}
U^{a}\hspace{0.1mm}_{b}U^{b}\hspace{0.1mm}_{c}=U^{a}\hspace{0.1mm}_{c}, \ \ \ U^{a}\hspace{0.1mm}_{a}=1 \ \ \ \ {\rm and} \ \ \ \ U^{a}\hspace{0.1mm}_{b}u^b=u^a
\eea
as well as 
\bea\label{rel_op_2}
h^a\hspace{0.1mm}_{b}h^b\hspace{0.1mm}_{c}=h^a\hspace{0.1mm}_{c},  \ \ \ \ h^a\hspace{0.1mm}_{a}=3 \ \ \ \ {\rm and} \ \ \ \ h^a\hspace{0.1mm}_{b}u^b=0
\eea
The space-time metric can be expressed as
\bea 
\label{metric} 
g_{ab}=h_{ab}+ u_{a}u_{b}
\eea
where $h_{ab}$ is the induced metric on the $3$-spaces orthogonal to $u^a$ at every point of space-time. Making use of the projection operators \eqref{projectiom_operators}, the covariant time derivative
\bea\label{time_cov_der}
\dot{A}^{a}\hspace{0.1mm}_{...b}:= u^{c}\nabla_{c}A^{a}\hspace{0.1mm}_{...b} 
\eea 
and the fully orthogonally projected covariant derivative 
\bea\label{space_cov_der} 
\bar{\nabla}_{c}A^{a}\hspace{0.1mm}_{...b}:= h^{d}\hspace{0.1mm}_{c}h^{a}\hspace{0.1mm}_{f}...h^{g}\hspace{0.1mm}_{b}\nabla_{d}A^{f}\hspace{0.1mm}_{...g} 
\eea
are defined for a generic tensor $A^{a}\hspace{0.1mm}_{...b}$. Moreover, given the Levi-Civita tensor $\varepsilon ^{abcd}$, we define the alternating tensor
\bea\label{3-Levi-Civita}
\varepsilon ^{abc}:=\varepsilon^{kabc}u_{k}\ \ \ \ {\rm with} \ \ \ \ \varepsilon ^{abc}u_c=0
\eea
According to \cite{cargese}, we denote by angle brackets the orthogonal projections of vectors $w^a$ and the orthogonally projected
symmetric trace-free part (PSTF) of tensors $A^{ab}$ of rank$\ =2$, namely
\bea\label{angle_brackets}
 w^{\langle a \rangle}:=h^{a}\hspace{0.1mm}_{b}w^{b} \ \ \ \ {\rm and} \ \ \ \ A^{\langle ab \rangle}:=\left[h^{(a}\hspace{0.1mm}_{c}h^{b)}\hspace{0.1mm}_{d}-\frac{1}{3}h^{ab}h_{cd} \right]A^{cd} 
\eea
Kinematical quantities are related to the splitting of the covariant derivative of the $4$-velocity $u^a$. We indeed have the following identity
\bea\label{cdu} 
\nabla_{a}u_{b}=\sigma_{ab}+\frac{1}{3}\Theta h_{ab}+\omega_{ab}+u_{a}\dot{u}_{b} 
\eea
where $\sigma_{ab}=\sigma_{\langle ab\rangle}:=\bar{\nabla}_{\langle a}u_{b\rangle}$ is the shear tensor, 
$\Theta:=\bar{\nabla}_{a}u^{a}$ is the expansion scalar, $\omega_{ab}=\omega_{[ab]}:=\bar{\nabla}_{[a}u_{b]}$ is the vorticity tensor and $\dot{u}_{b}=u^a\nabla_au_b$ is the acceleration vector. We also introduce the vorticity vector
\bea\label{vor} 
\omega^{a}:=\frac{1}{2}\varepsilon^{abc}\omega_{bc} \ \ \ \ \   \iff \ \ \ \ \  \omega_{ab}= - \varepsilon_{abc}\omega^{c} 
\eea
\\
and the magnitudes
\bea\label{s2o2} 
\sigma^2:=\frac{1}{2}\sigma^{ab}\sigma_{ab} \ \ \ \ {\rm and}  \ \ \ \ \omega^2:=\frac{1}{2}\omega^{ab}\omega_{ab} 
\eea
The Weyl conformal curvature tensor
\bea\label{Weyl2} 
C_{abcd} = R_{abcd} + \frac{1}{2}\left(g_{ad}R_{bc}-g_{ac}R_{bd}+g_{bc}R_{ad}-g_{bd}R_{ac}\right) + \frac{1}{6}R\left(g_{ac}g_{bd}-g_{ad}g_{bc}\right) 
\eea
written in terms of the Riemann curvature tensor $R_{abcd}$, the Ricci tensor $R_{ab}$ and the Ricci scalar $R$, can be decomposed by making use of the so called electric part $E_{ab}=E_{\langle ab \rangle}$ and magnetic part $H_{ab}=H_{\langle ab \rangle}$. The latter are defined as
\bea\label{EHWeyl} 
E_{ab}:=C_{cadb}u^{c}u^{d} \ \  \ \ {\rm and} \ \ \ \  H_{ab}:=\frac{1}{2}\varepsilon_{ade}C^{de}\hspace{0.1mm}_{bc}u^{c} 
\eea
from which we have the identity
\bea\nonumber\label{Weyl1} 
&&C^{abcd}=\left[4\delta^{[a}\hspace{0.1mm}_{k}\delta^{b]}\hspace{0.1mm}_{l}\delta^{[c}\hspace{0.1mm}_{n}\delta^{d]}\hspace{0.1mm}_{m}-\varepsilon^{ab}\hspace{0.1mm}_{kl}\varepsilon^{cd}\hspace{0.1mm}_{nm}\right]E^{kn}u^{l}u^{m}+\\
&& \ \ \ \ \ \ \ \ \ \ \ \ \ \ \ \ \ \ \ \ \ \  \ \ \ \ +2\left[\varepsilon^{ab}\hspace{0.1mm}_{kl}\delta^{[c}\hspace{0.1mm}_{n}\delta^{d]}\hspace{0.1mm}_{m}-\delta^{[a}\hspace{0.1mm}_{k}\delta^{b]}\hspace{0.1mm}_{l}\varepsilon^{cd}\hspace{0.1mm}_{nm}\right] H^{kn}u^{l}u^{m} 
\eea
Eqs. \eqref{Weyl1}, together with the expressions for the Ricci tensor and Ricci scalar derived from Einstein's equations, completely determine the Riemann curvature tensor $R_{abcd}$.

The energy--momentum tensor $T^{ab}=T^{(ab)}$ of matter is decomposed as 
\bea\label{SET} 
T_{ab}= \mu u_{a}u_{b} - ph_{ab}+ 2q_{(a}u_{b)}+\Pi_{ab} 
\eea 
where $\mu:=T_{ab}u^{a}u^{b}$ is the relativistic energy density, $p:=-\frac{1}{3}T_{ab}h^{ab}$ is the isotropic pressure, $q_{a}:= T_{bc}u^{b}h^{ca}$ is the relativistic momentum density and $\Pi_{ab}=\Pi_{\langle ab \rangle}:=T_{cd}h^{c}\hspace{0.1mm}_{\langle a}h^{d}\hspace{0.1mm}_{b \rangle}$ is a PSTF tensor which describes anisotropic pressure. For a perfect fluid, we have $q_{a}=0=\Pi_{ab}$. Possibly, the following energy conditions (respectively referred to as weak, dominant and strong energy conditions)
\begin{equation}\label{energy_conditions} 
\begin{cases}
\mu\geq 0 \\
\mu+p\geq 0
\end{cases}, \quad\quad \mu\geq |p|\quad, \quad\quad  
\begin{cases}
\mu+p\geq 0\\
\mu+3p\geq 0
\end{cases}
\end{equation}
can be required.

As for the field equations, a first set of six equations comes from suitable projections of the Ricci identities for the $4$-velocity $u^a$
\begin{eqnarray}\label{Ricci} 
\left(\nabla_{c}\nabla_{d}-\nabla_{d}\nabla_{c}\right)u_a=R_{abcd}u^b 
\end{eqnarray}
These first six equations are distinguished into propagation and constraint equations, respectively. In particular, we have:
\begin{itemize}
\item From $\left[\left(\nabla_{c}\nabla_{d}-\nabla_{d}\nabla_{c}\right)u_a - R_{abcd}u^b\right]u^{c}g^{da}=0$, we obtain the Raychaudhuri propagation equation
\begin{eqnarray}\label{Raychaudhuri1+3} 
\dot{\Theta}-\bar{\nabla}_{a}\dot{u}^{a}+\dot{u}_{a}\dot{u}^{a}+2\left(\sigma^2-\omega^2\right)+\frac{1}{3}\Theta^2+\frac{1}{2}\left(\mu+3p\right) = 0 
\end{eqnarray}

\item From $\left[\left(\nabla_{c}\nabla_{d}-\nabla_{d}\nabla_{c}\right)u_a - R_{abcd}u^b\right]u^{c}\varepsilon^{ade}=0$, we get the vorticity propagation equation
\begin{eqnarray}\label{dotvortic} 
\dot{\omega}^{\langle a\rangle} -\frac{1}{2}\varepsilon^{abc}\bar{\nabla}_{b}\dot{u}_{c}+\frac{2}{3} \Theta\omega^{a}-\sigma^{ad}\omega_{d}=0 
\end{eqnarray}

\item By applying the PSFT operator to $\left[\left(\nabla_{c}\nabla_{d}-\nabla_{d}\nabla_{c}\right)u_a - R_{abcd}u^b\right]u^{c}=0$, we deduce the shear propagation equation
\begin{eqnarray}\label{shearprop}
\dot{\sigma}_{\langle ad \rangle}+\dot{u}_{\langle a}\dot{u}_{d \rangle}-\bar{\nabla}_{\langle d}\dot{u}_{a \rangle}+\sigma_{\langle d}\hspace{0.1mm}^{c}\sigma_{a\rangle c}+\frac{2}{3}\Theta\sigma_{ad} -\omega_{\langle d}\hspace{0.1mm}\omega_{a \rangle} + E_{ad}+\frac{1}{2}\Pi_{ad}=0 
\end{eqnarray}

\item  From $\left[\left(\nabla_{c}\nabla_{d}-\nabla_{d}\nabla_{c}\right)u_a - R_{abcd}u^b\right]g^{ac}h^d\hspace{0.1mm}_{e}=0$, we obtain the constraint equation
\begin{eqnarray}\label{130alpha}
\bar{\nabla}^{a}\sigma_{da}+2\varepsilon_{dac}\dot{u}^a\omega^{c}-h_{de}\varepsilon^{eab}\bar{\nabla}_{a}\omega_{b}-\frac{2}{3}\bar{\nabla}_{d}\Theta-q_{d}=0 
\end{eqnarray}

\item From $\left[\left(\nabla_{c}\nabla_{d}-\nabla_{d}\nabla_{c}\right)u_a - R_{abcd}u^b\right]\varepsilon^{acd}=0$, we get the vorticity divergence identity
\begin{eqnarray}\label{vortdiver} 
\bar{\nabla}_{a}\omega^{a}+\dot{u}_{a}\omega^{a}=0 
\end{eqnarray}

\item The PSFT part of  $\left[\left(\nabla_{c}\nabla_{d}-\nabla_{d}\nabla_{c}\right)u_a - R_{abcd}u^b\right]\varepsilon^{ecd}=0$ yields the constraint
\begin{eqnarray}\label{hequation}  
H^{ca}=2\dot{u}^{\langle c}\omega^{a\rangle}+\varepsilon^{fd\langle c}\bar{\nabla}_{f}\sigma_{d}\hspace{0.1mm}^{a\rangle}-\bar{\nabla}^{\langle c}\omega^{a\rangle}  
\end{eqnarray}
\end{itemize}
A second set of field equations arises from the $(1+3)$-splitting of the conservation laws 
\bea \nabla_{b}T^{ab}=0 \eea
In detail:
\begin{itemize}
\item From $\nabla_{b}T^{ab}u_a=0$, we have
\begin{eqnarray}\label{eneconserv} 
\dot{\mu}+\Theta\left(\mu + p \right)-2\dot{u}^{a}q_{a}+\bar{\nabla}_{a}q^{a}-\Pi_{ab}\sigma^{ab}=0 
\end{eqnarray}

\item From $\nabla_{b}T^{ab}h^c\hspace{0.1mm}_{a}=0$, we obtain
\begin{eqnarray}\label{momconserv} 
\left(\mu + p \right)\dot{u}^{a}+\frac{4}{3}\Theta q^{a}+q^{b}\sigma_{b}\hspace{0.1mm}^{a} +q^{b}\omega_{b}\hspace{0.1mm}^{a}  +\dot{q}^{\langle a \rangle}-\bar{\nabla}^{a}p+\bar{\nabla}_{b}\Pi^{ab}-\Pi^{ab}\dot{u}_{b}=0  
\end{eqnarray}
\end{itemize}
A third set of equations is derived from the Kundt-Tr{\"u}mper equation
\bea\label{Bianchi} 
\nabla_{b}C_{ij}\hspace{0.1mm}^{ab}=-\nabla_{[i}\left[R^{a}\hspace{0.1mm}_{j]}-\frac{1}{6}R\delta^{a}\hspace{0.1mm}_{j]}\right] 
\eea
which is shown to be equivalent to the Bianchi identities $\nabla_{[a}R_{bc]de}=0$ \cite{Trumper}. In fact, by appropriately elaborating equation \eqref{Bianchi}, two propagation equations and two constraint equations for the electric and magnetic parts of the Weyl curvature tensor are obtained:
\begin{itemize}
\item By applying the PSTF operator to $\left[\nabla_{b}C_{ij}\hspace{0.1mm}^{ab}+\nabla_{[i}\left(R^{a}\hspace{0.1mm}_{j]}-\frac{1}{6}R\delta^{a}\hspace{0.1mm}_{j]}\right)\right]u^i=0$, we get the propagation equation
\begin{eqnarray}\label{pro_eq_E}
 \nonumber \dot{E}^{\langle i j \rangle}&=&-\Theta E^{\langle i j \rangle}+3E^{t \langle i}\sigma^{j\rangle}\hspace{0.1mm}_{t}-2\varepsilon^{t k\langle i}H_{k}\hspace{0.1mm}^{j\rangle}\dot{u}_{t}+E^{\sigma\langle i}\omega_{\sigma}\hspace{0.1mm}^{j\rangle}+\\
\nonumber&&+curl(H)^{ij}+\dot{u}^{\langle j}q^{i \rangle}+\frac{1}{2}\dot{\Pi}^{\langle i j \rangle}-\frac{1}{2}\bar{\nabla}^{\langle i}q^{j \rangle}-\frac{1}{2}\left(\mu+ p\right)\sigma^{ i j }+\\
&&\label{dote13}+\frac{1}{6}\Theta\Pi^{ij}+\frac{1}{2}\Pi^{k\langle j}\sigma^{i\rangle}\hspace{0.1mm}_{ k}+\frac{1}{2}\varepsilon^{t k \langle i}\Pi^{j \rangle}\hspace{0.1mm}_{k}\omega_{t} 
\end{eqnarray}
where $curl(H)^{ij}:=\varepsilon^{ab\langle i}\bar\nabla_a H^{j\rangle}\hspace{0.1mm}_b$.

\item By applying the PSTF operator to $\left[\nabla_{b}C_{ij}\hspace{0.1mm}^{ab}+\nabla_{[i}\left(R^{a}\hspace{0.1mm}_{j]}-\frac{1}{6}R\delta^{a}\hspace{0.1mm}_{j]}\right)\right]\varepsilon^{kij}=0$, we obtain the propagation equation
\begin{eqnarray}\label{pro_eq_H}
\nonumber\dot{H}^{\langle i j\rangle}&=&-curl(E)^{i j}-\frac{1}{2}curl\left(\Pi\right)^{i j}+2\varepsilon^{t k\langle i}E_{k}\hspace{0.1mm}^{j\rangle}\dot{u}_{t}-\Theta H^{\langle i j\rangle}+\\
&&\label{doth13}+3H^{\langle i}\hspace{0.1mm}_{k}\sigma^{j\rangle k}-\varepsilon^{kt\langle j}\omega_{k}H^{i\rangle}\hspace{0.1mm}_{t}-\frac{3}{2}q^{\langle i}\omega^{j\rangle}-\frac{1}{2}\varepsilon^{kt\langle i}q_{t}\sigma_{k}\hspace{0.1mm}^{j\rangle}
\end{eqnarray}
where $curl(E)^{ij}:=\varepsilon^{ab\langle i}\bar\nabla_a E^{j\rangle}\hspace{0.1mm}_b$ and $curl(\Pi)^{ij}:=\varepsilon^{ab\langle i}\bar\nabla_a \Pi^{j\rangle}\hspace{0.1mm}_b$.

\item  From $\left[\nabla_{b}C_{ij}\hspace{0.1mm}^{ab}+\nabla_{[i}\left(R^{a}\hspace{0.1mm}_{j]}-\frac{1}{6}R\delta^{a}\hspace{0.1mm}_{j]}\right)\right]u_au^ih^j\hspace{0.1mm}_k =0$, we deduce the constraint equation
\begin{eqnarray}\label{constr_eq_E}
\nonumber\bar{\nabla}_{j}\left(E_{i}\hspace{0.1mm}^{j} \right)&+&\varepsilon_{it k}H^{kj}\sigma_{j}\hspace{0.1mm}^{t}+3\omega_{\alpha}H^{\alpha}\hspace{0.1mm}_{i}=\\ 
&=&\label{econstr13}\frac{1}{3}\bar{\nabla}_{i}\mu+\frac{1}{2}\bar{\nabla}_{j}\left(\Pi_{i}\hspace{0.1mm}^{j} \right)+\frac{1}{3}
\Theta q_{i}-\frac{1}{2}q^{j}\sigma_{i j}-\frac{3}{2}\varepsilon_{ik j}\omega^{k}q^{j}
\end{eqnarray}

\item From $\left[\nabla_{b}C_{ij}\hspace{0.1mm}^{ab}+\nabla_{[i}\left(R^{a}\hspace{0.1mm}_{j]}-\frac{1}{6}R\delta^{a}\hspace{0.1mm}_{j]}\right)\right]u_a\varepsilon^{ijk}=0$, we derive the constraint equation
\begin{eqnarray}\label{contr_eq_H}
\nonumber\bar{\nabla}_{j}\left(H^{i j}\right)&-&3E^{i}\hspace{0.1mm}_{k}\omega^{k}-\varepsilon^{itk}E_{kj}\sigma^{j}\hspace{0.1mm}_{t}-\left(\mu+p\right)\omega^{i}-\\
&-&\label{hconstr13}\frac{1}{2}\varepsilon^{ikt}\bar{\nabla}_{k}q_{t}+\frac{1}{2}\varepsilon^{ikt}\sigma_{k j}\Pi_{t}\hspace{0.1mm}^{j}-\frac{1}{2}\omega^{t}\Pi_{t}\hspace{0.1mm}^{i}=0
\end{eqnarray}
\end{itemize}
The equations presented in this Subsection are the starting point for the further $(1+1+2)$-splitting, which is used in the following.
%%%%%%%%%%%%%%%%%%%%%%%%%%%%%%%%%%%%%%%%%%%%%%%%%%%%%%%%%%%%%%%%%%%%%%%%%%%
%%%%%%%%%%%%%%%%%%%%%%%%%%%%%%%%%%%%%%%%%%%%%%%%%%%%%%%%%%%%%%%%%%%%%%%%%%%
\subsection{$(1+1+2)$ covariant equations for LRS space-times in signature $(+---)$} \label{Subsection2.2}
The $(1+1+2)$-splitting relies on the introduction of an additional space-like vector field $n^{i}$, orthogonal to $u^i$, with $n^iu_i=0$ and $n^{i}n_{i}=-1$. At every point $x$ of space-time $M$, the $3$-space contained in $T_xM\/$ and orthogonal to $u^i$ is further decomposed into the direct sum of a $1$-dimensional subspace parallel to $n_{i}$ and a $2$-dimensional subspace orthogonal to both $u^i$ and $n^i$. As a consequence, the  metric tensor can be expressed as
\bea\label{mertric_1_1_2}
 g_{ij}=u_{i}u_{j}-n_{i}n_{j}+N_{ij} \ \ \ \ {\rm with} \ \ \ \ h_{ij}=-n_{i}n_{j}+N_{ij}
\eea
where $N_{ij}$ denotes the restriction of the metric tensor to the $2$-space orthogonal to $u^i$ and $n^i$. The tensor
\bea\label{projection_1_1_2}
N^i\hspace{0.1mm}_{j} =g^i\hspace{0.1mm}_{j} - u^{i}u_{j} + n^{i}n_{j}, \ \ \ {\rm with}  \ \ \ N^i\hspace{0.1mm}_{j}u^j=N^i\hspace{0.1mm}_{j}n^j =0,  \ \ \ N^i\hspace{0.1mm}_{j} N^j\hspace{0.1mm}_{h}=N^i\hspace{0.1mm}_{h}  \ \ \ {\rm and}  \ \ \ N^i\hspace{0.1mm}_{i}=2
\eea
plays the role of projection operator into the $2$-space orthogonal to $u^i$ and $n^i$. Therefore, every $3$-vector $V^{i}$ ($V^iu_i=0$) can be decomposed as
\bea\label{vector_decomposition_1_1_2} 
V^{i}=-Vn^{i}+\mathcal{V}^{i}
\eea
where $V=V^{i}n_{i}$ and $\mathcal{V}^{i}=N^{i}\hspace{0.1mm}_{j}V^{j}$. In an analogous way, every PSFT $3$-tensor $W_{ab}=W_{\langle ab\rangle}$ can be expressed as
\bea\label{tensor_decomposition_1_1_2} 
W_{ab}=W\left(n_{a}n_{b} + \frac{1}{2}N_{ab}\right) - 2\mathcal{W}_{(a}n_{b)} + \mathcal{W}_{ab} 
\eea
where $W=W^{ab}n_{a}n_{b}=W^{ab}N_{ab}$, $\mathcal{W}_{a}=N_{a}\hspace{0.1mm}^{b}n^{c}W_{bc}$ and $\mathcal{W}_{ab}=\left(N_{(a}\hspace{0.1mm}^{c}N_{b)}\hspace{0.1mm}^{d}-\frac{1}{2}N_{ab} N^{cd}\right)W_{cd}$. We also introduce the alternating tensor
\bea\label{2_alternating}
\varepsilon_{ab}:=\varepsilon_{jab}n^j=\varepsilon_{ijab}u^in^j \ \ \ \ {\rm with} \ \ \ \ \varepsilon_{ab}\varepsilon^{cd}= 
N_{a}\hspace{0.1mm}^{c}N_{b}\hspace{0.1mm}^{d} - N_{a}\hspace{0.1mm}^{d}N_{b}\hspace{0.1mm}^{c}
\eea
After that, making use of the fully orthogonally projected covariant derivative \eqref{space_cov_der}, we define two new derivatives given by
\bea\label{hat_derivative}
\hat{A}^a\hspace{0.1mm}_{...b}:=n^c\bar{\nabla}_c\/A^a\hspace{0.1mm}_{...b}
\eea
and
\bea\label{delta_derivative}
\delta_c\/A^a\hspace{0.1mm}_{...b}:= N_c\hspace{0.1mm}^{d}N^a\hspace{0.1mm}_{e}...N^f\hspace{0.1mm}_{b}\bar{\nabla}_d\/A^e\hspace{0.1mm}_{...f}
\eea
holding for every tensor $A^a\hspace{0.1mm}_{...b}$. The covariant derivatives of the $1$-forms $u_{a}$ and $n_{a}$ are expressed respectively as
\begin{subequations}\label{cdun0}
\begin{align}\label{cdu0}
\non \nabla_a\/u_b=& -u_a\left(A n_b-\mathcal{A}_b\right)+n_a n_b\left(\Sigma-\frac{1}{3} \Theta\right)  - n_{a}\left(\Sigma_{b}-\varepsilon_{b c}\Omega^c\right)-n_{b}\left(\Sigma_{a}+\varepsilon_{ac}\Omega^c\right) + \\
 & +  N_{a b}\left(\frac{1}{3} \Theta+\frac{1}{2} \Sigma\right)+\Omega \varepsilon_{ab}+\Sigma_{ab}
\end{align}
\be\label{cdn0}
\nabla_a\/n_b=  -Au_a\/u_b + u_a\alpha_b+\left(\Sigma-\frac{1}{3} \Theta\right)\/n_a u_b  -\left(\Sigma_a+\varepsilon_{ac} \Omega^c\right) u_b-n_a\/a_b+\frac{1}{2}\phi\/N_{ab}  +\xi\varepsilon_{ab}+\zeta_{ab}
\ee
\end{subequations}
where the scalar, vector, and tensor components appearing in eqs. \eqref{cdun0} are given by
\begin{subequations}\label{kinematical_quantities} 
\be
A:=\dot{u}_{i}n^i, \quad \Sigma:=\sigma_{ab}n^{a}n^{b}, \quad  \Omega:=\omega_{i}n^{i}, \quad \phi:=N^{ab}\delta_{a}n_{b}, \quad \xi:=\frac{1}{2}\varepsilon^{ab}\delta_{a}n_{b} 
\ee
\be
\mathcal{A}^a:=N^a{}_b\dot{u}^b, \quad  \alpha^a:= N^{a}{}_{b}{\dot{n}}^{b}, \quad a^c:=N^{c}{}_{b}\hat{n}^b, \quad \Sigma^a:=N^{ab}\sigma_{bc}n^c, \quad  \Omega^a:=N^a{}_b\omega^b
\ee
\be
\Sigma_{ab}:=\left(N_{(a}{}^{c}N_{b)}{}^{d}\!-\!\frac{1}{2}N_{ab}N^{cd}\right)\sigma_{cd},\quad \zeta_{ab}:=\left(N_{(a}{}^{c}N_{b)}{}^{d}\!-\!\frac{1}{2}N_{ab}N^{cd}\right)\delta_c\/n_d
\ee
\end{subequations}
In Locally Rotationally Symmetric (LRS) geometries, at each point of space-time the unit vector $n^i$ indicates a preferred spatial direction, coinciding with a local axis of symmetry. All observations are identical under rotations about $n^i$, that is, observations are the same in all spatial directions perpendicular to $n^i$. In particular, this implies that all tensors having physical meaning must have null projections into the $2$-space orthogonal to both $u^i$ and $n^i$. As a consequence, in an LRS space-time, the covariant derivatives of $u_{i}$ and $n_{i}$ reduce to
\bea \label{cdubis} 
&& \nabla_{i}u_{j}=\Sigma\left(n_{i}n_{j}+\frac{1}{2}N_{ij}\right)+\frac{1}{3}\Theta\left(N_{ij}-n_{i}n_{j}\right)-Au_{i}n_{j}+\Omega\varepsilon_{ij} \\
\label{cdn} && \nabla_{i}n_{j}=\frac{1}{2}\phi N_{ij}+\xi\varepsilon_{ij}-Au_{i}u_{j}+\left(\Sigma-\frac{1}{3}\Theta \right)n_{i}u_{j} 
\eea
Moreover, denoting by
\bea\label{E_H}
&& E:=E_{ab}n^{a}n^{b} \ \ \ \ {\rm and} \ \ \ \ H:=H_{ab}n^{a}n^{b} 
\eea 
the following identities necessarily hold
\begin{subequations}\label{decompositions}
\begin{align}
\dot{u}^{a}&=-An^{a}\\
\omega^{a}&=-\Omega n^{a}\\
\sigma_{ab}&=\Sigma\left(n_{a}n_{b}+\frac{1}{2}N_{ab}\right)\\
E_{ab}&=E\left(n_{a}n_{b}+\frac{1}{2}N_{ab}\right)\\
H_{ab}&=H\left(n_{a}n_{b}+\frac{1}{2}N_{ab}\right)
\end{align}
\end{subequations}
Similarly, denoting by
\bea\label{Q_PI}
Q:=q^in_i \ \ \ \ {\rm and} \ \ \ \ \Pi:=\Pi_{ij}n^in^j
\eea
we have the following representations
\bea\label{Q_PI_bis}
q_i=-Qn_i \ \ \ \ {\rm and} \ \ \ \ \Pi_{ij}=\Pi\left(n_{i}n_{j}+\frac{1}{2}N_{ij}\right)
\eea
for the momentum density vector and the anisotropic pressure tensor, respectively. In the variables above, the weak and strong energy conditions take the form 
\bea\label{energy_conditions_bis}
\begin{cases} 
\mu\geq 0\\
\mu+p+\Pi\geq 0
\end{cases} \ , \quad\quad  
\begin{cases}
\mu+3p\geq 0 \\
\mu+p+\Pi\geq 0
\end{cases}
\eea
Summing it all up, the variables that covariantly describe LRS space-times are the scalar quantities
\bea\label{scalar_quantities}
\{A,\Theta,\Sigma,\Omega,\phi,\xi,E,H,\mu,p,Q,\Pi\}
\eea
For these variables, corresponding equations are then needed. Such equations, typically divided into evolution, propagation, evolution-propagation, and constraint equations, are partly derived from the $(1+3)$-equations obtained in Subsection 2.1, and partly by working out the Ricci identities for the vector field $n^i$. 
\\
\\
{\it Evolution equations}:
\begin{itemize}
\item By saturating eq. \eqref{dotvortic} with $n_a$, we get
\bea\label{dotomega} 
\dot{\Omega} = -A\xi-\frac{2}{3}\Theta\Omega-\Omega\Sigma
\eea
\item By saturating eq. \eqref{doth13} with $n_in_j$, we obtain
\bea\label{dot_H} 
\dot{H} = 3E\xi+\frac{3}{2}\Pi\xi-\Theta H -\frac{3}{2}H\Sigma -\Omega Q 
\eea
\item From the identity $\left[\left(\nabla_{c}\nabla_{d}-\nabla_{d}\nabla_{c}\right)n_a - R_{abcd}n^b\right]u^cN^{ad}=0$, we have
\bea\label{dot_phi}
\dot{\phi} = - \left(\Sigma+\frac{2}{3}\Theta\right)\left(A+\frac{1}{2}\phi\right) + 2\Omega\xi - Q 
\eea
\item From the identity $\left[\left(\nabla_{c}\nabla_{d}-\nabla_{d}\nabla_{c}\right)n_a - R_{abcd}n^b\right]u^{c}\varepsilon^{ad}=0$, we deduce
\bea\label{dot_xi}
\dot{\xi} = -\left(\frac{1}{2}\Sigma+\frac{1}{3}\Theta\right)\xi - \Omega\left(\frac{1}{2}\phi+A\right) + \frac{1}{2}H 
\eea
\end{itemize}

\noindent
{\it Propagation equations}:
\begin{itemize}
\item eq. \eqref{vortdiver} amounts to
\bea\label{hat_Omega}
\hat{\Omega} = - \Omega\left(A+\phi\right)
\eea
\item The identity $\left[\left(\nabla_{c}\nabla_{d}-\nabla_{d}\nabla_{c}\right)n_a - R_{abcd}n^b\right]n^{c}N^{ad}=0$ yields directly
\bea\label{hat_phi} 
\hat{\phi} = -\frac{1}{2}\phi^{2} + 2\xi^{2} - \left(\Sigma-\frac{1}{3}\Theta\right)\left(\Sigma+\frac{2}{3}\Theta\right) - \frac{2}{3}\mu -\frac{1}{2}\Pi + E 
\eea
\item The identity $\left[\left(\nabla_{c}\nabla_{d}-\nabla_{d}\nabla_{c}\right)n_a - R_{abcd}n^b\right]n^{c}\varepsilon^{ad}=0$ gives rise to 
\bea\label{hatxi} 
\hat{\xi} = - \left(\Sigma-\frac{1}{3}\Theta\right)\Omega - \phi\xi 
\eea
\item By saturating eq. \eqref{130alpha} with $n^d$, we get 
\bea\label{hat_Theta_Sigma} 
\frac{2}{3}\hat{\Theta} + \hat{\Sigma}= -\frac{3}{2}\Sigma\phi + 2\Omega\xi - Q
\eea
\item By saturating \eqref{econstr13} with $n^i$, we have
\bea\label{hat_E} 
\hat{E}-\frac{1}{2}\hat{\Pi}+\frac{1}{3}\hat{\mu} = - \frac{3}{2}\phi\left(E-\frac{1}{2}\Pi\right) - 3\Omega H - Q\left(\frac{1}{3}\Theta+\frac{1}{2}\Sigma\right)
\eea
\item By saturating \eqref{hconstr13} with $n_i$, we obtain
\bea\label{hat_H} 
\hat{H} = - \frac{3}{2}\phi H - \Omega\left(-3E+\mu+p-\frac{1}{2}\Pi\right) + Q\xi
\eea
\end{itemize}

\noindent
{\it Evolution--Propagation equations}:
\begin{itemize}
\item By working out the Raychaudhuri equation \eqref{Raychaudhuri1+3}, we get
\bea\label{Raychaudhuri1+1+2}  
\dot{\Theta} + \hat{A} = - A\phi + A^{2} - \frac{1}{3}\Theta^{2} - \frac{3}{2}\Sigma^{2} + 2\Omega^{2} - \frac{1}{2}\left(\mu+3p\right) 
\eea
\item From eq. \eqref{eneconserv}, we obtain
\bea\label{dot_mu} 
\dot{\mu} - \hat{Q}= - \left(\mu + p\right)\Theta + Q\phi - 2AQ +\frac{3}{2}\Sigma\Pi 
\eea 
\item By saturating eq. \eqref{momconserv} with $n_a$, we have 
\bea\label{dot_Q} 
\dot{Q}-\hat{\Pi} -\hat{p} = + \frac{3}{2}\Pi\phi - A\left(\mu + p + \Pi\right) - \frac{4}{3}Q\Theta + Q\Sigma
\eea
\item By saturating eq. \eqref{shearprop} with $n^an^d$, we get
\bea\label{dot_Sigma} 
\dot{\Sigma}-\frac{2}{3}\hat{A} = - \frac{1}{3}A\phi + \frac{1}{2}\Sigma^{2} + \frac{2}{3}\Omega^{2} - \frac{2}{3}A^{2} - \frac{2}{3}\Theta\Sigma - E - \frac{1}{2}\Pi 
\eea
\item By saturating eq. \eqref{dote13} with $n_in_j$, we deduce 
\bea\label{dot_E} 
\dot{E} - \frac{1}{2}\dot{\Pi} + \frac{1}{3}\hat{Q} = -E\Theta - \frac{3}{2}E\Sigma - 3H\xi + \frac{2}{3}AQ + \frac{1}{6}Q\phi - \frac{1}{2}\left(\mu+p\right)\Sigma + \frac{1}{6}\Theta\Pi - \frac{1}{4}\Pi\Sigma 
\eea
\item The identity $\left[\left(\nabla_{c}\nabla_{d}-\nabla_{d}\nabla_{c}\right)n_a - R_{abcd}n^b\right]u^{c}n^{d}u^{a}=0$ provides
\bea\label{equazione_indipendente} 
\hat{A} - \dot{\Sigma} + \frac{1}{3}\dot{\Theta} = - \left(\Sigma-\frac{1}{3}\Theta\right)^{2} + A^{2} - \frac{1}{6}\left(\mu+3p\right) + \frac{1}{2}\Pi + E 
\eea
Eq. \eqref{equazione_indipendente} is dependent on the previous ones. Indeed, it is given by the linear combination $\frac{1}{3}\eqref{Raychaudhuri1+1+2} - \eqref{dot_Sigma}$. 
\end{itemize}

\noindent
{\it Constraint equation}:
\begin{itemize}
\item Saturating eq. \eqref{hequation} with $n_cn_a$, we get
\bea\label{constraint_H} 
H=2A\Omega+\Omega\phi-3\xi\Sigma
\eea
which expresses the magnetic part of the Weyl tensor in terms of other kinematical variables.
\end{itemize}
%%%%%%%%%%%%%%%%%%%%%%%%%%%%%%%%%%%%%%%%%%%%%%%%%%%%%%%%%%%%%%%%%%%%%%%%
%%%%%%%%%%%%%%%%%%%%%%%%%%%%%%%%%%%%%%%%%%%%%%%%%%%%%%%%%%%%%%%%%%%%%%%%
\section{Consistency of the covariant equations} \label{Section3}
Following the lines drawn in \cite{Ellisperf,Ellisnonperf}, we discuss the conditions that ensure consistency and integrability of the covariant equations presented in the previous section. To this end, we preliminarily observe that in LRS space-times every covariantly defined scalar quantity $f$ must satisfy the relation
\be\label{rif}
\dot{f}\Omega=\hat{f}\xi
\ee
Indeed, the requirement that the spatial derivatives, in the directions perpendicular to $n^i$, must be zero ($\delta_{a}f=0$) implies the identity
\be\label{gradiente_covariante}
\nabla_i\/f = \dot{f}u_i - \hat{f}n_i
\ee
and then
\be\label{derivata_seconda}
\nabla_j\nabla_i\/f = \nabla_j\left(\dot{f}\right)u_i + \dot{f}\nabla_j\/u_i - \nabla_j\left(\hat{f}\right)n_i - \hat{f}\nabla_j\/n_i
\ee
Relation \eqref{rif} is then deduced by saturating eq. \eqref{derivata_seconda} with $\varepsilon^{ji}$. Eq. \eqref{rif} provides us with a consistency condition for the covariant equations. Another useful identity is the commutation relation for the {\it dot}  and {\it hat} derivatives of a scalar function $f$
\bea\label{comts} 
\dot{\hat f} - \hat{\dot f} = -A\dot{f}+\Sigma\hat{f}-\frac{1}{3}\Theta\hat{f} 
\eea
which helps us to investigate the integrability of the covariant equations.

By applying identity \eqref{rif} for $f=\Omega$ and $f=\xi$ separately, and using the evolution and propagation equations for $\Omega$ and 
$\xi$ (eqs. \eqref{dotomega}, \eqref{dot_xi}, \eqref{hat_Omega} and \eqref{hatxi}), as well as the expression \eqref{constraint_H} for $H$, we obtain the following set of two equations
\bea
\begin{cases}
\left(\frac{2}{3}\Theta\Omega + \Sigma\Omega - \phi\xi\right)\Omega=0 \\
\left(\frac{2}{3}\Theta\Omega + \Sigma\Omega - \phi\xi\right)\xi=0 
\end{cases}
\eea
which necessarily implies that the relation 
\bea\label{rif1} 
\phi\xi=\left(\Sigma+\frac{2}{3}\Theta\right)\Omega 
\eea
must always hold.

Another constraint is deduced by imposing the commutation relation \eqref{comts} for $f=E$. In fact, from eqs. \eqref{hat_E} and \eqref{dot_E} we can derive the expressions for $\hat E$ and $\dot E$ respectively. Then, substituting them into \eqref{comts}, using the covariant equations as well as the constraint \eqref{rif1}, we get the equation 
\bea\label{rif2}
\left(p+\mu+\Pi\right)\xi\Omega=Q\left(\Omega^{2}+\xi^{2}\right) 
\eea
which has to be satisfied. In particular, from \eqref{rif2} we infer that
\bea\label{condizione_Q=0} 
Q=0 \qquad \iff \qquad \left(p+\mu+\Pi\right)\xi\Omega=0  
\eea
Applying the same procedure to the quantities $\{\Omega,\xi,\phi,\Sigma+\frac{2}{3}\Theta\}$ does not result in any further constraint. The same holds true when requiring the vanishing of the {\it dot} derivative of the constraint \eqref{constraint_H}.  

Instead, the integrability condition \eqref{comts} for $H$ and the compatibility between the expression \eqref{constraint_H} of $H$ and the propagation equation \eqref{hat_H} give rise to two conditions that need to be discussed. More in detail, deriving $\dot H$ and $\hat H$ from \eqref{dot_H} and \eqref{hat_H} and substituting into \eqref{comts}, we obtain the relation
\bea\nonumber && \frac{1}{6} \biggl[\left(18E+6\mu+6p+6\Pi\right)\Sigma+\left(12E-2p-2\mu-2\Pi\right)\Theta-12AQ+6\dot{p}+\\
 \label{rif3} &&  \ \ \ \ \ \ \ \ \ \ \ \ \ \ \ \ \ \ \   +6\hat{Q}-12\dot{\Pi}\biggr]\Omega=\left[Q\Sigma+\frac{2}{3}Q\Theta+3\phi E-\frac{3}{2}\phi\Pi+\hat{p}+\hat{\mu}-2\hat{\Pi}\right]\xi 
\eea
At the same time, replacing eq. \eqref{hat_H} into the {\it hat} derivative of eq. \eqref{constraint_H}, we have   
\bea  
\nonumber && \Omega \biggl[-2A^{2}+2\Sigma^{2}-\frac{4}{3}\Theta\Sigma+\frac{2}{9}\Theta^2-4\xi^2-2E+2\hat{A}+\\
\label{rif4} && \ \ \ \  \ \ \ \ \ \ \ \ \ \ \ \ \ \ \ \ \ +\frac{1}{3}\mu+p-\Pi\biggr]=-3\xi\left[\Sigma\phi+\frac{2}{3}\hat{\Theta}+\frac{2}{3}Q\right] 
\eea
Therefore, consistency of the covariant equations requires that conditions (\ref{rif}), (\ref{rif1}), (\ref{rif2}), (\ref{rif3}), and (\ref{rif4}) must be satisfied. In order to discuss these conditions, it is convenient to distinguish two main cases: $\Omega\xi=0$ and $\Omega\xi\neq0$. In the following, we will classify LRS space-times of types I, II, and III based on the requirement $\Omega\xi=0$, regardless of the type of perfect or non-perfect fluid. This is different from the original definition given in \cite{Ellisperf}, but it will be more useful for our purposes.
%%%%%%%%%%%%%%%%%%%%%%%%%%%%%%%%%%%%%%%%%%%%%%%%%%%%%%%%%%%%%%%%%%%%%
\subsection{The case $\Omega\xi=0$}\label{Subsection_3.1}
In this circumstance, we can point out three different subcases:
\begin{enumerate}
\item {\bf LRS space-times of class I}: $\Omega\neq0$ and $\xi=0$.
\\
From eq. \eqref{rif}, it follows that $\dot f=0$ for every covariantly defined scalar function $f$. Then, from eqs. \eqref{hatxi}, \eqref{rif1} and \eqref{rif2}, we have $Q=\Sigma=\Theta=0$. In this case, the constraint \eqref{rif3} is automatically verified, whereas the condition \eqref{rif4} is satisfied in view of eq. \eqref{equazione_indipendente}. Also, note that all the evolution equations become trivial identities.
\item {\bf LRS space-times of class II}: $\Omega=0$ and $\xi=0$. Under such conditions, all the constraints \eqref{rif}, \eqref{rif1}, \eqref{rif2}, \eqref{rif3} and \eqref{rif4} are automatically satisfied.

\item {\bf LRS space-times of class III}: $\Omega=0$ and $\xi\neq0$.
\\
In this case, eq. \eqref{rif} implies $\hat f=0$ for every covariantly defined scalar function $f$. Moreover, from eqs. \eqref{rif1} and \eqref{rif2}, we get $\phi=0$ and $Q=0$ respectively. In addition, from the evolution equation \eqref{dotomega}, we have $A=0$. As a consequence, equations \eqref{rif3} and \eqref{rif4} are identically verified.
\end{enumerate}
%%%%%%%%%%%%%%%%%%%%%%%%%%%%%%%%%%%%%%%%%%%%%%%%%%%%%%%%%%%%%%%%%%%%%%%%%%%%%%%
\subsection{The case $\Omega\xi\neq 0$}\label{subsection3.2}
In this case, equation \eqref{rif} creates a constraint between the evolution and propagation equations. For instance, inserting the content of the evolution and propagation equations for $\phi$ (eqs. \eqref{dot_phi} and \eqref{hat_phi}) into eq. \eqref{rif}, we can derive the following expression for $E$.
\bea\label{espressione_E1}
E=-2\xi^2+2\Omega^2+\frac{1}{3}\Theta\Sigma+\Sigma^2-\frac{2}{9}\Theta^2+\frac{2}{3}\mu+\frac{1}{2}\Pi-\frac{A\Omega}{\xi}\left(\Sigma+\frac{2}{3}\Theta\right)-\frac{Q\Omega}{\xi} 
\eea
A direct calculation shows that the equation \eqref{espressione_E1}, together with eqs. \eqref{rif1} and \eqref{rif2}, makes the constraints \eqref{rif3} and \eqref{rif4} automatically satisfied.  Moreover, making use of eqs. \eqref{constraint_H}, \eqref{rif1}, \eqref{rif2} and \eqref{espressione_E1}, it is easily seen that the constraint \eqref{rif} holds identically for $f\in\{\Sigma+\frac{2}{3}\Theta,E,H,\Omega,\xi\}$ too. Finally, we notice that eqs. \eqref{constraint_H}, \eqref{rif1}, \eqref{rif2} and \eqref{espressione_E1} allow us to express the quantities $\{H,E,\phi,Q\}$ as functions of the remaining variables.
%%%%%%%%%%%%%%%%%%%%%%%%%%%%%%%%%%%%%%%%%%%%%%%%%%%%%%%%%%%%%%%%%%%%%%%%%%%%%%%%
%%%%%%%%%%%%%%%%%%%%%%%%%%%%%%%%%%%%%%%%%%%%%%%%%%%%%%%%%%%%%%%%%%%%%%%%%%%%%%%%

\section{Polar formalism}\label{Section4}
\subsection{Spinor fields in polar form and $(1+1+2)$ covariant decomposition}
In this section, after briefly reviewing the main features of the polar formalism for spinor fields \cite{Fabbri:2023onb, Fabbri:2024avj, Fabbri:2023dgv}, we implement the $(1+1+2)$-decomposition of such polar formulation.

To begin, given a set of Clifford matrices $\gamma^\mu$ ($\mu=0,\ldots,3$) and a tetrad field $e^i_{\mu}$ (with dual co-tetrad $e^\mu_i$, $e^i_\mu\/e^\nu_i=\delta^\nu_\mu$, $e^i_\mu\/e^\mu_j=\delta^i_j$), we denote by $\boldsymbol{\gamma}^i:=\gamma^\mu\/e_\mu^i$ and by
$\boldsymbol{s}_{ik}\!:=\![\boldsymbol{\gamma}_{i},\boldsymbol{\gamma}_{k}]/4$ the corresponding generators of the complex Lorentz group. The parity-odd $\gamma^{5}$ matrix is implicitly defined through the relation
\begin{eqnarray}
&2i\boldsymbol{s}_{ab}\!=\!\varepsilon_{abcd}\gamma^{5}\boldsymbol{s}^{cd}
\end{eqnarray}
Given a spinor field $\psi$, its adjoint spinor is defined as $\bar{\psi}\!=\!\psi^{\dagger}\gamma^{0}$. In this paper, we will work with regular spinors, defined by the requirement that either $i\bar{\psi}\gamma^{5}\psi\neq0$ or $\bar{\psi}\psi\neq0$ be always verified. Regular spinors can always be written in the so-called polar form, which, in chiral representation, reads
\begin{eqnarray}
&\psi\!=\!\sqrt{\rho/2}e^{-\frac{i}{2}\beta\gamma^{5}}
\ \boldsymbol{L}^{-1}\left(\begin{tabular}{c}
$1$\\
$0$\\
$1$\\
$0$
\end{tabular}\right)
\label{spinor}
\end{eqnarray}
where the functions $\rho$ and $\beta$ are called {\it density} and {\it chiral angle} and where $\boldsymbol{L}$ has the structure of a spinor transformation \cite{jl1, jl2}. The functions $\rho$ and $\beta$ satisfy the relations
\begin{eqnarray}
&i\bar{\psi}\gamma^{5}\psi\!=\!\rho\sin{\beta} \quad {\rm and} \quad \bar{\psi}\psi\!=\!\rho\cos{\beta}\label{scalars}
\end{eqnarray}
in such a way that the remaining independent bilinears can be expressed as
\begin{eqnarray}
&\bar{\psi}\boldsymbol{\gamma}^{a}\gamma^{5}\psi\!=\!\rho s^{a} \quad {\rm and} \quad \bar{\psi}\boldsymbol{\gamma}^{a}\psi\!=\!\rho u^{a}\label{vectors}
\end{eqnarray}
where the unit vector fields $u^a$ and $s^a$ meet the conditions (see, for example, Appendix B in \cite{Fabbri_proof})
\begin{equation}
u_{a}u^{a}\!=\!-s_{a}s^{a}\!=\!1 \quad {\rm and} \quad u_{a}s^{a}\!=\!0
\end{equation}
Therefore, the two vector fields $u^{a}$ and $s^{a}$ identify with the $4$-vector fields of the velocity and the spin axial vector fields, respectively. Their mutual orthogonality makes them natural candidates to represent the unit tangent vectors to the time-like and space-like congruences needed in the $(1+1+2)$ covariant decomposition. In fact, with the identification $n^i=s^i$, the Dirac field naturally provides the basic elements for the $(1+1+2)$-splitting. Systematically assuming the identification $n^i=s^i$, we can then apply the geometrical framework illustrated in Section \ref{Section2} in order to formulate a covariant $(1+1+2)$-approach to the Dirac field.

Now, in polar formulation, the covariant derivative of a spinor field can always be expressed in the form \cite{Fabbri:2023dgv}
\begin{equation}\label{decspinder}
\boldsymbol{\nabla}_{k}\psi\!=\!(\frac{1}{2}\nabla_{k}\ln{\rho}\mathbb{I}
\!-\!\frac{i}{2}\nabla_{k}\beta\gamma^{5}
\!-\!iP_{k}\mathbb{I}
\!-\!\frac{1}{2}R_{abk}\boldsymbol{s}^{ab})\psi
\end{equation}
where the tensors $P_{k}$ and $R_{abk}\!=\!-R_{bak}$, respectively referred to as \emph{momentum} and \emph{tensorial connection}, are involved. In the particular case of plane waves in flat space-time, eq. (\ref{decspinder}) would result into $i\boldsymbol{\nabla}_{k}\psi\!=\!P_{k}\psi$ so that $P_{k}$ would be precisely the momentum of the plane wave. The tensorial connection is related to the velocity and the spin $4$-vector fields by the identities
\begin{equation}\label{identità_R}
\nabla_{k}s_{b}\!=\!s^{a}R_{abk} \quad {\rm and} \quad \nabla_{k}u_{b}\!=\!u^{a}R_{abk}
\end{equation}
as was proven in \cite{Fabbri:2023dgv}. Making use of eqs. \eqref{identità_R} and defining $\varepsilon_{ab}=\varepsilon_{abcd}u^cs^d$, we can get the following expression for the tensorial connection
\begin{eqnarray}\label{Rfull}
R_{abm}\!=\!u_{a}\nabla_{m}u_{b}\!-\!u_{b}\nabla_{m}u_{a}
\!+\!s_{b}\nabla_{m}s_{a}\!-\!s_{a}\nabla_{m}s_{b}
\!+\!(u_{a}s_{b}-u_{b}s_{a})s^{k}\nabla_{m}u_{k}
\!+\!2\varepsilon_{ab}V_{m}
\end{eqnarray}
Eq. \eqref{Rfull} describes the tensorial connection in terms of the covariant derivatives of spin and velocity, as well as a further vector field $V_{m}$. The presence of the vector field $V_{m}$ shows that the covariant derivatives of spin and velocity cannot encode all the remaining information about the spinor field in addition to the quantities $\rho$ and $\beta$. In fact, let us consider, for instance, the spinor field in its rest frame with spin aligned along the third axis: in this case $\boldsymbol{L}\!=\!\mathbb{I}$ in \eqref{spinor}. In this frame, rotations around the third axis cannot affect the velocity (whose spatial components are zero) and, by construction, the spin or their covariant derivatives. On the other hand, these rotations do have an impact on the spinor field, and they must be encoded within the covariant derivative of the spinor field itself. This means that rotations around the spin axis must be encoded either in $P_{m}$ or in $V_{m}$, which is the part of $R_{abm}$ not given by the covariant derivatives of velocity and spin. Furthermore, only the difference $P_{m}\!-\!V_{m}$ has physical significance and, because $P_{m}$ is the momentum of the matter distribution, $P_{m}\!-\!V_{m}$ has to be recognized as the {\it effective momentum}.

Eventually, by using eqs. \eqref{cdun0}, the tensorial connection \eqref{Rfull} can be written as
\begin{equation}
\begin{split}\label{Rfull1+1+2}
R_{ab}{}^{m}
=&-2A u^m u_{[a} s_{b]}-2\left(\Sigma-\frac{1}{3} \Theta\right) s^m s_{[a}u_{b]}-2s^{m}u_{[a}\left(\Sigma_{b]}-\varepsilon_{b]c} \Omega^c\right) - 2u_{[a} s_{b]}\Sigma^m\\
& +2u^m u_{[a}\mathcal{A}_{b]} -2 u^{m} s_{[a}\alpha_{b]} + 2s^{m} s_{[a}a_{b]} + 2u_{[a}N^{m}{}_{b]}\left(\frac{1}{3} \Theta+\frac{1}{2} \Sigma\right)- \phi s_{[a} N^{m}{}_{b]} \\
& +2u_{[a}\Sigma^{m}{}_{ b]} -2s_{[a}\zeta^{m}{}_{b]}+2\Omega u_{[a}\varepsilon^{m}{}_{b]}-2\xi s_{[a}\varepsilon^{m}{}_{b]}
- 2u_{[a} s_{b]}\varepsilon^{m}{}_{c}\Omega^c + 2\varepsilon_{ab}V^{m}
\end{split}
\end{equation}
About the dynamical character, we have that the Dirac equation
\begin{equation}\label{D0}
i\boldsymbol{\gamma}^{j}\boldsymbol{\nabla}_{j}\psi\!-\!m\psi\!=\!0
\end{equation}
can also be polarly decomposed. The first step is to substitute in it the decomposition (\ref{decspinder}), getting
\begin{eqnarray}
\left[(\nabla_{a}\beta\!+\!B_{a})\boldsymbol{\gamma}^{a}\gamma^5
\!+\!i(\nabla_{a}\ln{\phi^{2}}\!+\!R_{a})\boldsymbol{\gamma}^{a}
\!+\!2P_{a}\boldsymbol{\gamma}^{a}\!-\!2m\mathbb{I}\right]\psi\!=\!0\label{D}
\end{eqnarray}
in which the identity $\boldsymbol{\gamma}_{i}\boldsymbol{\gamma}_{j}\boldsymbol{\gamma}_{k}\!
=\!\boldsymbol{\gamma}_{i}\eta_{jk}\!-\!\boldsymbol{\gamma}_{j}\eta_{ik}\!+\!\boldsymbol{\gamma}_{k}\eta_{ij}\!+\!i\varepsilon_{ijkq}\gamma^5\boldsymbol{\gamma}^{q}$ was used and where the notations $R_{a}\!:=\!R_{ab}^{\phantom{ab}b}$ and $B_{a}\!:=\!\frac{1}{2}\varepsilon_{abcd}R^{bcd}$ were introduced. As second step, we multiply eq. \eqref{D} on the left, in turn, by $\bar{\psi}\boldsymbol{\gamma}^{i}$ and $\bar{\psi}\boldsymbol{\gamma}^{i}\gamma^5$, each time splitting imaginary and real parts, obtaining
\be
\bar{\psi}\boldsymbol{\gamma}^{j}: \begin{cases}
\mathrm{Im}\ \longrightarrow & \nabla_{j}\left(\bar{\psi}\psi\right)
\!-\!iB_{j}\bar{\psi}\gamma^5\psi\!+\!R_{j}\bar{\psi}\psi\!+\!4iP^{k}\bar{\psi}\boldsymbol{s}_{kj}\psi\!=0\\
\mathrm{Re}\ \longrightarrow & 2i\nabla_{k}\left(\bar{\psi}\boldsymbol{s}^{kj}\psi\right)
\!+\!iR^{pqj}\bar{\psi}\boldsymbol{s}_{pq}\psi\!-\!2P^{j}\bar{\psi}\psi
\!+\!2m\bar{\psi}\boldsymbol{\gamma}^{j}\psi\!=\!0
\end{cases}\label{forwardmomentum}
\ee
\be
\bar{\psi}\boldsymbol{\gamma}^{j}\gamma^5: \begin{cases}
\mathrm{Im}\ \longrightarrow & 2\nabla^{k}\left(\bar{\psi}\boldsymbol{s}_{kj}\gamma^5\psi\right)
\!+\!R_{pqj}\bar{\psi}\boldsymbol{s}^{pq}\gamma^5\psi\!+\!2iP_{j}\bar{\psi}\gamma^5\psi\!=\!0\\
\mathrm{Re}\ \longrightarrow & i\nabla_{j}\left(\bar{\psi}\gamma^5\psi\right)\!+\!B_{j}\bar{\psi}\psi
\!+\!iR_{j}\bar{\psi}\gamma^5\psi\!-\!4P^{k}\bar{\psi}\boldsymbol{s}_{kj}\gamma^5\psi\!+\!2m\bar{\psi}\boldsymbol{\gamma}_j\gamma^5\psi\!=\!0
\end{cases}\label{complementarymomentum}
\ee
called Gordon decompositions \cite{Fabbri:2024avj}. By expressing all bilinears in polar form and substituting them in the real part of \eqref{forwardmomentum} and in the imaginary part of \eqref{complementarymomentum}, we obtain, after some calculations, the following equations
\begin{subequations}\label{Dirac_decomposed0}
\be
u^{j}(R_{j}\!+\!\nabla_{j}\ln{\rho})\!=\!0\label{d1} 
\ee
\be
s^{j}(R_{j}\!+\!\nabla_{j}\ln{\rho})\!=\!2m\sin{\beta}
\ee
\be
(B_{j}\!+\!\nabla_{j}\beta)\varepsilon^{jk}\!=\!0
\ee
\be
P^{k}\!=\!m\cos{\beta}u^{k}\!+\!(B_{j}\!+\!\nabla_{j}\beta)u^{[j}s^{k]}
\!+\!\frac{1}{2}(R_{j}\!+\!\nabla_{j}\ln{\rho})\varepsilon^{jk}\label{momentum}
\ee
\end{subequations}
These last equations can then be substituted in \eqref{D} to prove that the Dirac equation is actually verified \cite{Fabbri:2019tad}. As a consequence, eqs. \eqref{Dirac_decomposed0} are equivalent to the Dirac equation \eqref{D0}. Making use of eq. \eqref{Rfull1+1+2} and performing all the indicated projections, eqs. \eqref{Dirac_decomposed0} assume the final form
\begin{subequations}\label{Dirac_decomposed}
\be
\Theta\!+\!\dot{(\ln{\rho})}\!=\!0
\ee
\be
\phi\!-\!A\!+\!\hat{(\ln{\rho})}\!-\!2m\sin{\beta}\!=\!0
\ee
\be
\alpha^k\!\varepsilon_{ka}-\!2\Omega_{a}\!+\!\delta_a\beta\!=\!0
\ee
\be
2(P\!-\!V)_{i}u^{i}\!=\!2m\cos{\beta}\!-\!2\Omega\!-\!\hat{\beta}\label{Pu}
\ee
\be
2(P\!-\!V)_{i}s^{i}\!=-\!2\xi\!-\!\dot{\beta}\label{Ps}
\ee
\be
2(P\!-\!V)_{i}N^{ik}\!=(a_{j}\!-\mathcal{A}_{j}\!+\!\delta_{j}\ln{\rho})\varepsilon^{jk}\label{PN}
\ee
\end{subequations}
in which we see that only the difference $(P\!-\!V)_{i}$ is dynamically significant. This is the reason why only the effective momentum $P_{m}\!-\!V_{m}$ is physically meaningful.
%%%%%%%%%%%%%%%%%%%%%%%%%%%%%%%%%%%%%%%%%%%%%%%%%%%%%%%%%%%%%%%%%%%%%%%%%%%%%%%%%%%%%%%%%%%%%%%%%%%
%%%%%%%%%%%%%%%%%%%%%%%%%%%%%%%%%%%%%%%%%%%%%%%%%%%%%%%%%%%%%%%%%%%%%%%%%%%%%%%%%%%%%%%%%%%%%%%%%%%
\subsection{The energy--momentum tensor for the hydrodynamic representation of the spinor field}\label{V}
We provide a representation of the energy--momentum tensor of the spinor field, suitable for a hydrodynamic description of the spinor field itself. To this end, let 
\be\label{DEMT}
T^{rs}\!=\!
\frac{i}{8}(\bar{\psi}\boldsymbol{\gamma}^{r}\boldsymbol{\nabla}^{s}\psi
\!-\!\boldsymbol{\nabla}^{s}\bar{\psi}\boldsymbol{\gamma}^{r}\psi
\!+\!\bar{\psi}\boldsymbol{\gamma}^{s}\boldsymbol{\nabla}^{r}\psi
\!-\!\boldsymbol{\nabla}^{r}\bar{\psi}\boldsymbol{\gamma}^{s}\psi)
\ee
be the usual form of the energy--momentum tensor of the Dirac field. We notice that our convention for Einstein equations is $G_{ij}=T_{ij}$, where $G_{ij}$ denotes the Einstein tensor and $T_{ij}$ the energy--momentum tensor.  Compared to the convention $G_{ij}=\frac{1}{2}T_{ij}$, used by other authors, we include an additional factor $\frac{1}{2}$ in the definition of the energy--momentum tensor. Due to this, all the quantities arising from the energy--momentum tensor \eqref{DEMT} contain this additional factor. Another remark is about the terminology we use. Following one of the conventions in literature, we call \eqref{DEMT} energy--momentum tensor. However, as we will see, \eqref{DEMT} is unrelated to the momentum $P^i$ introduced in the previous Section.

That said, by making use of the expression \eqref{decspinder}, the spinor energy--momentum tensor \eqref{DEMT} can be rewritten in the form
\be\label{energy}
T^{rs}\!=\!\frac{1}{4}\rho\!\left(P^{s}u^{r}+P^{r}u^{s}
+\nabla^{r}\beta s^{s}/2
\!+\!\nabla^{s}\beta s^{r}/2
\!-\!\frac{1}{4}R_{a n}^{\phantom{a n}s}\varepsilon^{r a n m}s_{m}
\!-\!\frac{1}{4}R_{a n}^{\phantom{a n}r}\varepsilon^{s a n m}s_{m}\right)
\ee
After that, by inserting the expression of the tensorial connection \eqref{Rfull1+1+2} as well as that of the effective momentum \eqref{Pu}, \eqref{Ps}, and \eqref{PN}, from eq. \eqref{energy} we get the following representation for the spinor energy--momentum tensor
\be\label{T1+1+2}
\begin{split}
T^{rs}
= &\frac{1}{4}\rho\!\left[\left(2m\cos{\beta}\!-\!2\Omega\!-\!\hat{\beta}\right)u^{r}u^{s}+2\left(\xi\!+\!\dot{\beta}\right)s^{(r}u^{s)}-\hat{\beta}s^{(r}s^{s)}+\Omega N^{rs}\right.\\
&\left.
+2\left(\mathcal{A}_{c}-a_{j}\!-\!\delta_{j}\ln{\rho}\right)\varepsilon^{j (r}u^{s)}-2\left(\Sigma_{n}+\varepsilon_{nc} \Omega^c+\varepsilon^{s) c}\delta^{c}\beta\right)s^{(r}\varepsilon^{s) n}-2\varepsilon^{n (r }\Sigma^{s)}{}_{ n}\right]
\end{split}
\ee
On the other hand, in the $(1+1+2)$ framework a generic energy--momentum tensor can be decomposed as
\be\label{T}
T_{ab}\!=\!\mu u_{a}u_{b}\!-\!p(N_{ab}\!-\!s_{a}s_{b})\!-\!Q(s_{a}u_{b}\!+\!s_{b}u_{a})\!+\!(Q_{a}u_{b}\!+\!Q_{b}u_{a})
\!+\!\frac{1}{2}\Pi(N_{ab}\!+\!2s_{a}s_{b})
\!+\!(\Pi_{a}s_{b}\!+\!\Pi_{b}s_{a})\!+\!\Pi_{ab}
\ee
in terms of the projections
\begin{subequations}\label{componenti_TEI}
\be
\mu\!=\!T_{ab}u^{a}u^{b}\label{energy-density}
\ee
\be
p\!=\!-\frac{1}{3}T_{ab}(N^{ab}\!-\!s^{a}s^{b})
\ee
\be
Q\!=\!T_{ab}s^{a}u^{b}
\ee
\be
\Pi\!=\!\frac{1}{3}T_{ab}(N^{ab}\!+\!2s^{a}s^{b})
\ee
\be
Q^{a}\!=\!T_{cd}N^{ca}u^{d}
\ee
\be
\Pi^{a}\!=\!-T_{cd}N^{ca}s^{d}
\ee
\be
\Pi^{ab}\!=\!\left(N^{ac}N^{bd}\!-\!\frac{1}{2}N^{ab}N^{cd}\right)T_{cd}
\ee
\end{subequations}
In view of this, by applying the projection procedure \eqref{componenti_TEI} to the tensor \eqref{T1+1+2}, we end up with the quantities
\begin{subequations}\label{thermoquantities}
\be\label{MuPsi}
\mu\!=\!\frac{\rho}{2}\left(m\cos{\beta}\!-\!\frac{\hat{\beta}}{2}-\!\Omega\!\right) 
\ee
\be\label{thermo_p}
p\!=\!-\frac{1}{12}\rho\left(\hat{\beta}\!+\!2\Omega\right)
\ee
\be\label{thermo_Pi}
\Pi\!=\!-\frac{1}{6}\rho\left(\hat{\beta}\!-\!\Omega\right)
\ee
\be\label{thermo_Q}
Q\!=\!-\frac{1}{4}\rho\left(\dot{\beta}\!+\!\xi\right)
\ee
\be\label{2_Q}
Q^{a}\!=-\frac{1}{4}\rho\varepsilon^{j a}\left(\delta_{j}\ln{\rho}-2\mathcal{A}_{j}+a_{j}\!\right)
\ee
\be\label{2_Pi}
\Pi^{a}\!=\frac{1}{4}\rho\left(\delta^a\beta+\Sigma_{n}\varepsilon^{s a}+ \Omega^a\right)
\ee
\be\label{anisopresstens}
\Pi^{ab}=-\frac{\rho}{4}\Sigma^{(a}_{j}\varepsilon^{b)j}
\ee
\end{subequations}
which represent the components of the spinor energy--momentum tensor expressed in hydrodynamic form. The quantities \eqref{thermoquantities} are expressed in terms of the fundamental variables of the $(1+1+2)$ covariant formalism, together with the density $\rho$ and the chiral angle $\beta$ of the spinor field. More information on the thermodynamical properties of the effective fluid described by the energy momentum tensor \eqref{T1+1+2} is given in \cite{Fabbri:2025ffi}.
%%%%%%%%%%%%%%%%%%%%%%%%%%%%%%%%%%%%%%%%%%%%%%%%%%%%%%%%%%%%%%%%%%%%%%%%%
%%%%%%%%%%%%%%%%%%%%%%%%%%%%%%%%%%%%%%%%%%%%%%%%%%%%%%%%%%%%%%%%%%%%%%%%%
\section{Spinorial fluid in LRS space-times}\label{Section5}
In this section, we present a first attempt at a covariant approach to the Dirac field, without making use of the tetrad formalism. The idea is to perform a matching between the covariant $(1+1+2)$-splitting and the polar formalism that we described in the previous Sections.

In the following, we will focus exclusively on LRS space-times, choosing the time-like vector field $u^i$ to coincide with the velocity $4$-vector field of the Dirac field, and the space-like vector field $n^i$ to coincide with the spin $4$-vector field $s^i$. Our construction is based on the hydrodynamic description of the Dirac field that we gave in Section \ref{Section4}. In particular, in accordance with the symmetries of the LRS geometry, the $2$-spatial quantities \eqref{2_Q}, \eqref{2_Pi}, and \eqref{anisopresstens} must be set equal to zero. In such a circumstance, the energy--momentum tensor of the Dirac field reduces to
\be\label{EMTLRS}
T_{ab}\!=\!\mu u_{a}u_{b}\!-\!p(N_{ab}\!-\!s_{a}s_{b})\!-\!Q(s_{a}u_{b}\!+\!s_{b}u_{a})\!+\!\frac{1}{2}\Pi(N_{ab}\!+\!2s_{a}s_{b})
\ee  
where the quantities $\mu$, $p$, $Q$ and $\Pi$ are given by eqs. \eqref{MuPsi}-\eqref{thermo_Q}. Moreover, the vanishing of the vector fields $Q^a$ and $\Pi^a$ implies that both the density $\rho$ and the chiral angle $\beta$ have to be covariantly defined, i.e., $\delta_i\rho=0$ and $\delta_i\beta=0$.

The $(1+1+2)$-covariant equations, discussed in Section \ref{Section2}, will now be coupled with the Dirac equations \eqref{Dirac_decomposed}. In connection with this, we notice that the Dirac equations \eqref{Pu}, \eqref{Ps} and \eqref{PN} have already been used to deduce the expression of the effective momentum $P^i-V^i$ and to obtain the expression of the hydrodynamic quantities (\ref{thermoquantities}a)-(\ref{thermoquantities}d) in terms of $\rho$, $\beta$, $\Omega$ and $\xi$. The remaining Dirac equations (\ref{Dirac_decomposed}a)-\ref{Dirac_decomposed}c) reduce to
\bea\label{diracequa}
\begin{cases}
\dot{\ln{\rho}} = - \Theta \\
\hat{\ln{\rho}} =2m\sin{\beta} + A - \phi
\end{cases}
\eea
Eqs. \eqref{diracequa} give us information about the evolution of the density $\rho$ along the time-like and space-like congruences. The analogous information regarding the chiral angle $\beta$ will be deduced from the conservation laws \eqref{dot_mu} and \eqref{dot_Q}, which in turn must be true since the Dirac equations imply them.

As for eqs. \eqref{diracequa}, we need to discuss their consistency and integrability. To this end, by applying eq. \eqref{rif} for $f=\ln{\rho}$ and using eqs. \eqref{diracequa}, we obtain the relation
\bea\label{vincA} 
\dot{(\ln{\rho})}\Omega=\hat{(\ln{\rho})}\xi \quad \Longleftrightarrow \quad  -\Theta\Omega=\left(2m\sin{\beta}+A-\phi\right)\xi 
\eea 
At the same time, the Dirac equations \eqref{diracequa} and the commutation relations \eqref{comts} for $f=\ln\rho$ yield the equation
\bea\label{dotA}
\dot{A}= \non &-& 2m\dot{\beta}\cos{\beta} - \left(\Sigma+\frac{2}{3}\Theta\right)\left(A+\frac{1}{2}\phi\right) + 2\Omega\xi - Q - \hat{\Theta} +\\
 &+& A\Theta+\left(\Sigma-\frac{1}{3}\Theta\right)\left(2m\sin{\beta}+A-\phi\right) 
\eea
Eqs. \eqref{vincA} and \eqref{dotA} provide us with the consistency and integrability conditions for the Dirac equations \eqref{diracequa}. In particular, if $\xi \not =0$, the following expression 
\bea\label{vincA1} 
A=-\frac{\Theta\Omega}{\xi}+\phi-2m\sin{\beta}  
\eea
for $A$ is derived. In this connection, a direct check shows that eqs. \eqref{dotA} and \eqref{vincA1} are consistent. The {\it dot} derivative $\dot A$, obtained from eq. \eqref{vincA1} and substituted into equation \eqref{dotA}, makes eq. \eqref{dotA} an identity. In addition to this, evaluating the {\it hat} derivative of eq. \eqref{vincA1}, it is seen that $\dot{A}\Omega=\hat{A}\xi$ if and only if $\dot{\beta}\Omega=\hat{\beta}\xi$. The latter condition must be verified by the derivatives of the chiral angle $\beta$.

In the following subsections, we will discuss the coupling with the Dirac field under two different assumptions: the case where the spinorial fluid is perfect and the non--perfect case.

%%%%%%%%%%%%%%%%%%%%%%%%%%%%%%%%%%%%%%%%%%%%%%%%%%%%%%%%%%%%%%%%%%%%%%%%%%%%%%%%%%%%%%%
%%%%%%%%%%%%%%%%%%%%%%%%%%%%%%%%%%%%%%%%%%%%%%%%%%%%%%%%%%%%%%%%%%%%%%%%%%%%%%%%%%%%%%%
\subsection{Perfect spinorial fluid}
Assuming that the spinorial fluid is perfect means imposing both the momentum density and the anisotropic pressure equal to zero, i.e., $q^i=-Qs^i=0$ and $\Pi^{ij}=\Pi\left(s^is^j+\frac{1}{2}N^{ij}\right)=0$. In this regard, directly from the expressions \eqref{thermo_Pi} and \eqref{thermo_Q}, we have 
\begin{subequations}\label{condizioni_fluido_perfetto}
\begin{align}
\label{condQ} q_i=0 \ \ &\iff \ \ \dot{\beta}=-\xi \\
\label{condPi}\Pi_{ij}=0  \ \ &\iff \ \ \hat{\beta}=\Omega 
\end{align}
\end{subequations}
Thus, when the Dirac field is seen to behave like a perfect fluid, the {\it dot} and {\it hat} derivatives of the chiral angle are directly connected to twist and vorticity, respectively. However, by applying the kinematic constraint \eqref{rif} for $f=\beta$, from the equations \eqref{condizioni_fluido_perfetto} we get the relation
\bea\label{perfluiconst} 
\dot{\beta}\Omega=\hat{\beta}\xi \ \ \Longrightarrow -\xi\Omega=\Omega\xi \ \ \iff \ \ \Omega\xi=0 
\eea
The conclusion follows that, when the time-like and space-like congruences of the $(1+1+2)$-splitting coincide with those of the $4$-velocity and $4$-spin of the Dirac field, a perfect spinorial fluid is only compatible with LRS space-times of class I, II, or III. Let's analyze the three different scenarios in detail.
\\
\\
{\bf LRSI}: $\xi=0$ and $\Omega \not =0$.
\\ 
As we have already seen in Section \ref{Section3}, in this case, we have 
\bea\label{condizioni_LRSI}
\Sigma=\Theta=0 \ \ {\rm and} \ \ \dot{f}=0  \ \ \forall f \ \text{ covariant scalar}   
\eea
The constraint \eqref{dotA} is automatically verified, whereas the covariant equations that are not identically satisfied are:
\begin{subequations}
\begin{align}
& \label{1}A\phi +\hat{A}- A^{2}-2\Omega^{2}+\frac{1}{2}\left(\mu+3p\right)=0 \\
% &\label{2}\dot{\mu}=0 \ ; \\
&  \label{3}\hat{p} - A\left(\mu+ p\right)=0 \\
& \label{4}-\frac{2}{3}\hat{A}+\frac{1}{3}A\phi-\frac{2}{3}\Omega^{2}+\frac{2}{3}A^{2} + E=0 \\
& \label{7} \hat{\Omega}+\Omega\left(A+\phi\right)=0  \\
& \label{8} \hat{\phi}+\frac{1}{2}\phi^{2}+\frac{2}{3}\mu-E=0  \\
&  \label{9}\hat{E}+\frac{1}{3}\hat{\mu}+\frac{3}{2}\phi E+3\Omega H =0  \\
& \label{10} \hat{H}+\frac{3}{2}\phi H+\Omega\left(-3E+\mu+p\right)=0  \\
& \label{12} \hat{\ln{\rho}}=2m\sin{\beta} + A - \phi
\end{align}
\end{subequations}
with now
\begin{subequations}
\begin{align}
&\label{115} \mu=\frac{1}{2}\rho\left[m\cos\beta-\frac{3}{2}\Omega\right] \\
&\label{116} p=-\frac{1}{4}\rho\Omega 
\end{align}
\end{subequations}
By combining the equation \eqref{1} with \eqref{4}, we obtain 
\bea\label{evin} 
E=-A\phi+2\Omega^2-\frac{1}{3}\left(\mu+3p\right) 
\eea
Substitution of eq. \eqref{evin} into eqs. \eqref{9} and \eqref{10} make them identically satisfied. So, we remain with the following set of six differential equations
\bea\label{lrs1s} 
\begin{cases}
\label{111} \hat{A}=-A\phi+ A^{2}+2\Omega^{2}-\frac{1}{4}\rho\left(m\cos\beta-3\Omega\right) \\
\label{113} \hat{\Omega}=-\Omega\left(A+\phi\right)  \\
\label{114} \hat{\phi}=-\frac{1}{2}\phi^{2}-A\phi+2\Omega^2-\frac{1}{2}\rho\left(m\cos\beta-2\Omega\right)  \\
\label{117} \hat{\ln{\rho}}=2m\sin{\beta}+A-\phi \\
\label{118} \hat{\beta}=\Omega \\
\label{112} \frac{1}{2}\hat{\left(\rho\Omega\right)}= - A\rho\left(m\cos\beta-2\Omega\right)
\end{cases} 
\eea
for five unknowns $\{A,\Omega,\phi,\beta,\rho\}$. Given appropriate initial data, the first five equations \eqref{lrs1s} can be solved uniquely for all the unknowns. The remaining sixth and independent equation is therefore a stringent constraint on the solutions that would thus be found. More in detail, by working it out, we obtain the relation
\be\label{lrs1sbis}
\phi = m\sin\beta + \frac{A}{\rho\Omega}\left(m\rho\cos\beta - 2\rho\Omega\right)
\ee
which is not automatically preserved along the solutions of the first five equations \eqref{lrs1s}. This means that the dynamics \eqref{lrs1s} is not (everywhere) tangent to the submanifold \eqref{lrs1sbis}. A constraint algorithm has to be applied here. A first step identifies a further submanifold of \eqref{lrs1sbis} described by 
\begin{equation}\label{lrs1sbis2}
\rho=\frac{\left(2A^2m^2 - 2\Omega^2\/m^2\right)\cos^2\beta + 12m\Omega\left(\frac{Am\sin\beta}{3} + A^2 + \Omega^2\right)\cos\beta -24\Omega^4 + \left(-24A^2 + 2m^2\right)\Omega^2}{\Omega\left(m\cos\beta-5\Omega\right)\left(m\cos\beta-2\Omega\right)}
\end{equation}
Requiring that the dynamics preserves both the constraints \eqref{lrs1sbis} and \eqref{lrs1sbis2} yields an additional submanifold. The latter is described by a Cartesian equation (here omitted for brevity) for the remaining variables $A$, $\Omega$, and $\beta$, which can not be explicitly solved for any of its variables. Therefore, we are not able to proceed further through the constraint algorithm. The conclusion follows that certainly there are no solutions in which all variables $(A,\Omega,\phi,\rho,\beta)$ are free. The constraint algorithm does not stabilize after the first two steps, so at most only two variables would remain free. However, due to purely computational reasons, we are not able to establish whether the problem admits solutions or not.
\\
\\
{\bf LRSII}: $\xi=\Omega=0$
\\ 
A first consequence of $\xi=\Omega=0$ (see eq. \eqref{constraint_H}) is
\bea\label{H_LRSII}
H=0
\eea
The equations for $\dot{H}$ and $\hat{H}$ (eqs. \eqref{dot_H} and \eqref{hat_H}) are identically satisfied. Also the evolution and propagation equations for $\Omega$ and $\xi$ (eqs. \eqref{dotomega}, \eqref{dot_xi}, \eqref{hat_Omega} and \eqref{hatxi}) are automatically verified. So, we are left with the equations concerning the remaining quantities $\Theta,\Sigma,A,\phi,E,\rho,\beta$. In this regard, the perfect fluid assumption gives us the condition
\bea\label{condizioni_beta_LRSII}
\begin{cases}
\dot{\beta}=-\xi=0\\
\hat{\beta}=\Omega=0
\end{cases}
\Longrightarrow \beta=\text{constant}
\eea
which implies
\bea
\begin{cases}
\mu=\frac{1}{2}\rho\/m\cos{\beta}\\
p=0
\end{cases}
\eea
Therefore, in LRSII space-times, the perfect spinorial fluid is necessarily a dust. Moreover, assuming $\mu\not =0$, the vanishing of the pressure $p=0$ implies $A=0$ (see eq. \eqref{dot_Q}). The worldlines of the time-like congruence are then geodesics. We also notice that the Dirac equation $\dot{\ln\rho}=-\Theta$  entails the conservation law \eqref{dot_mu} for the energy density. In fact, we have the identity
\bea 
\dot{\mu}=\frac{1}{2}\dot{\rho}m\cos{\beta}=-\frac{1}{2}\rho\/m\Theta\cos{\beta}=-\mu\Theta 
\eea
The covariant equations for the remaining undetermined variables are
\begin{subequations}\label{fieleq}
\be\label{dot_theta_LRSII}
\dot{\Theta}+\frac{1}{3}\Theta^{2}+\frac{3}{2}\Sigma^{2}+\frac{1}{2}\mu=0
\ee
\be\label{dot_Sigma_LRSII}
\dot{\Sigma}-\frac{1}{2}\Sigma^{2}+\frac{2}{3}\Theta\Sigma+E=0
\ee
\be\label{dot_phi_LRSII}
\dot{\phi}+\frac{1}{2}\phi\left(\Sigma+\frac{2}{3}\Theta\right)=0
\ee
\be\label{dot_E_LRSII}
\dot{E} + \Theta\/E + \frac{3}{2}E\Sigma + \frac{1}{2}\mu\Sigma=0 
\ee
\be\label{hat_Sigma_LRSII}
\hat{\Sigma}+\frac{2}{3}\hat{\Theta}+\frac{3}{2}\Sigma\phi=0
\ee
\be\label{hat_E_LRSII}
\hat{E}+\frac{1}{3}\hat{\mu}+\frac{3}{2}E\phi=0
\ee
\be\label{hat_phi_LRSII}
\hat{\phi}+\frac{1}{2}\phi^{2}+\left(\Sigma-\frac{1}{3}\Theta\right)\left(\Sigma+\frac{2}{3}\Theta\right)+\frac{2}{3}\mu-E=0
\ee
\be\label{dot_lnrho_LRSII}
\dot{\ln{\rho}} + \Theta =0
\ee
\be\label{hat_lnrho_LRSII}
\hat{\ln{\rho}} - 2m\sin{\beta} + \phi =0
\ee
\end{subequations}
In addition to eqs. \eqref{fieleq}, the constraint \eqref{dotA} has to be imposed too. In particular, the constraint \eqref{dotA} assumes here the explicit form
\be\label{constraint_101}
\hat\Theta=-\frac{3\Sigma\phi}{2} + 2\left(\Sigma - \frac{\Theta}{3}\right)m\sin\beta
\ee 
providing us with a propagation equation for $\Theta$. Eq. \eqref{constraint_101} represents a crucial difference compared to the covariant equations for spatially inhomogeneous LRSII dust models \eqref{dot_theta_LRSII}-\eqref{dot_lnrho_LRSII}, for which a general solution algorithm exists \cite{Ellisperf}. Indeed, now consistency between eqs. \eqref{dot_theta_LRSII} and \eqref{constraint_101} has to be imposed. In this regard, the commutation relations \eqref{comts} together with eqs. \eqref{dot_theta_LRSII} and \eqref{constraint_101} yield the relation
\be\label{integrability_ThetaLRSII}
4m\left(-\frac{m\rho\cos\beta}{3} + 2\Sigma^2 -\frac{4\Theta\Sigma}{3} + \frac{2\Theta^2}{9} + E\right)\sin\beta -3\left(E-\frac{m\rho\cos\beta}{6}\right)\phi=0
\ee
which expresses the integrability condition for eqs. \eqref{dot_theta_LRSII} and \eqref{constraint_101} (note that eq. \eqref{integrability_ThetaLRSII} simplifies dramatically for $\beta=k\pi$). 
The search for general solutions of the system of equations \eqref{fieleq}  and \eqref{integrability_ThetaLRSII} deserves a dedicated investigation and goes beyond the scope of the present work. Here, borrowing ideas again from \cite{Ellisperf}, we analyze some particular cases which admit solution under appropriate simplifying hypotheses: 1) $E=0$, 2) $\Sigma=0$, or 3) $\phi=0$. 
\\
\\
1) $E=0$.
\\
Continuing to suppose $\mu\not =0$, from eqs. \eqref{condizioni_beta_LRSII}, \eqref{dot_E_LRSII} and \eqref{hat_E_LRSII} we obtain
\bea\begin{cases}\label{Sigma_rho_LRSII}
\Sigma=0\\
\hat{\rho}=0
\end{cases}
\eea
Inserting the content of eq. \eqref{Sigma_rho_LRSII} into eqs. \eqref{fieleq}, we get the further conditions
\bea\begin{cases}
\phi=2m\sin\beta\\
\hat\Theta =0\\
\phi\Theta=0
\end{cases}
\eea
which, together with $A=\Sigma=0$, satisfy the constraint \eqref{dotA}. We can therefore distinguish two distinct subcases: 1.a) $\phi=0$ and 1.b) $\Theta=0$.
\\
\\
1.a) $\phi=0 \Rightarrow \beta=k\pi$. So we are left with only two variables $\Theta$ and $\rho$, and the set of equations
\bea\label{condizioni_1.a_LRSII} 
\begin{cases}
\frac{1}{3}\Theta^2 - \mu=0 \\
\dot{\Theta}+\frac{1}{2}\Theta^2=0 \\
\dot{\ln{\rho}} + \Theta=0
\end{cases}
\eea
representing a spatially flat FLRW space-time, filled with a spinorial dust. We will discuss the solution of \eqref{condizioni_1.a_LRSII} in Section 6.
\\
\\
1.b) $\Theta=0$. This subcase is only admissible under the very special condition $\mu=0$. In this circumstance, the energy--momentum tensor of the Dirac field is zero, even though the Dirac field is not zero. We necessarily have $\beta=\frac{\pi}{2}+k\pi$. The only two remaining variables are $\phi$ and $\rho$, which must satisfy the equations
\bea\label{condizioni_1.b_LRSII} 
\begin{cases}
\dot{\phi}=0\\
\dot{\rho}=0\\
\hat{\phi}+\frac{1}{2}\phi^2=0\\
\hat\ln{\rho}=2m\sin{\beta}-\phi
\end{cases}
\eea
Solutions of the system \eqref{condizioni_1.b_LRSII} exist and will be discussed in Section 6. They describe a very particular spinor field with a vanishing energy-momentum tensor, filling a flat space-time. In coordinates, solutions of this kind have already been found in \cite{Critical_solutions}. Although critical from a physical point of view, such solutions are allowed by mathematics.
\\
\\
2) $\Sigma=0$.
\\
From eq. \eqref{dot_Sigma_LRSII}, we deduce immediately that $E=0$. So we fall back into case 1), which we have already discussed above.
\\
\\
3) $\phi=0$.
\\
The evolution equation \eqref{dot_phi_LRSII} is identically satisfied. The propagation equation \eqref{hat_phi_LRSII} yields the expression
\bea \label{efi0} 
E=\left(\Sigma-\frac{1}{3}\Theta\right)\left(\Sigma+\frac{2}{3}\Theta\right)+\frac{2}{3}\mu 
\eea
for $E$. The remaining covariant equations assume the form
\bea\label{equaz_cov_phi=0_LRSII} \begin{cases}
\dot{\Theta}+\frac{1}{3}\Theta^{2}+\frac{3}{2}\Sigma^{2}+\frac{1}{2}\mu=0\\
\dot{\Sigma}-\frac{1}{2}\Sigma^{2}+\frac{2}{3}\Theta\Sigma+\left(\Sigma-\frac{1}{3}\Theta\right)\left(\Sigma+\frac{2}{3}\Theta\right)+\frac{2}{3}\mu=0\\
\dot{\ln{\rho}} + \Theta=0\\
\hat{\ln{\rho}}=2m\sin{\beta}\\
\hat{\Sigma}+\frac{2}{3}\hat{\Theta}=0\\
\hat{E}+\frac{1}{3}\hat{\mu}=0\\
\dot{E}=-\Theta E-\frac{3}{2}E\Sigma-\frac{1}{2}\mu\Sigma 
\end{cases}\eea
The consistency between expression \eqref{efi0} and the evolution and propagation equations for $E$ must be imposed. In this connection, a direct check shows that the evolution equation for $E$ is automatically verified. Instead, inserting eq. \eqref{efi0} into the propagation equation for $E$, we get the equation 
\bea\label{condizione_hat_E_LRSII} 
\hat{\Theta}\left(\Sigma+\frac{2}{3}\Theta\right) - \rho m^2\sin{\beta}\cos{\beta}=0 
\eea
We end up with three evolution equations and three propagation equations 
\bea\label{equaz_cov_phi=0_LRSII_final} 
\begin{cases}
\dot{\Theta}+\frac{1}{3}\Theta^{2}+\frac{3}{2}\Sigma^{2}+\frac{1}{2}\mu=0\\
\dot{\Sigma}-\frac{1}{2}\Sigma^{2}+\frac{2}{3}\Theta\Sigma+\left(\Sigma-\frac{1}{3}\Theta\right)\left(\Sigma+\frac{2}{3}\Theta\right)+\frac{2}{3}\mu=0\\
\dot{\ln{\rho}} + \Theta=0\\
\hat{\ln{\rho}}=2m\sin{\beta}\\
\hat{\Sigma}+\frac{2}{3}\hat{\Theta}=0\\
\hat{\Theta}\left(\Sigma+\frac{2}{3}\Theta\right) - \rho m^2\sin{\beta}\cos{\beta}=0 
\end{cases}
\eea
for the variables $\Theta$, $\Sigma$ and $\rho$. But the equations \eqref{equaz_cov_phi=0_LRSII_final} must be coupled to the constraint \eqref{constraint_101} which now reads as
\be\label{dotA_phi=0_LRSII}
\hat\Theta = 2m\sin\beta\/\left(\Sigma -\frac{1}{3}\Theta\right)
\ee
Despite this additional condition, the problem admits solutions. For instance, by requiring homogeneity ($\hat\Theta=\hat\Sigma=\hat\rho=0$), eqs. \eqref{equaz_cov_phi=0_LRSII_final} and \eqref{dotA_phi=0_LRSII} reduce to the system
\bea\label{sistema_finale_phi=0_LRSII} 
\begin{cases}
\dot{\Theta}+\frac{1}{3}\Theta^{2}+\frac{3}{2}\Sigma^{2}+\frac{1}{2}\mu=0\\
\dot{\Sigma}-\frac{1}{2}\Sigma^{2}+\frac{2}{3}\Theta\Sigma+\left(\Sigma-\frac{1}{3}\Theta\right)\left(\Sigma+\frac{2}{3}\Theta\right)+\frac{2}{3}\mu=0\\
\dot{\ln{\rho}} + \Theta=0
\end{cases}
\eea
with $\beta=k\pi$ and $\mu=\frac{1}{2}\rho\/m\cos\beta$. As we will see in Section 6, the system \eqref{sistema_finale_phi=0_LRSII} describes a Bianchi-I space-time, filled with a spinorial dust.
\\
\\
{ \bf LRSIII}: $\xi\not =0$ and $\Omega=0$
\\
In this case, from Section \ref{Section3} we have
\bea\label{condizioni_LRSIII}
\Omega=\phi=0 \ \ \text{and} \ \ \hat{f}=0 \ \  \forall f \ \text{covariant scalar} 
\eea
From the evolution equation for $\Omega$, eq \eqref{dotomega}, we get immediately
\bea 
A\xi=0 \ \Longrightarrow \ A=0 
\eea
Thus, from the Dirac equations \eqref{diracequa} and due to the condition $\hat{\ln{\rho}}=0$, we have the condition
\bea 
2m\sin{\beta}=0 \ \Longrightarrow \ \beta=k\pi
\eea
This implies $\xi=-\dot{\beta}=0$, which contradicts the LRSIII assumption. Thus, in the case of a perfect spinorial fluid, no LRSIII solution exists.
%%%%%%%%%%%%%%%%%%%%%%%%%%%%%%%%%%%%%%%%%%%%%%%%%%%%%%%%%%%%%%%%%%%%%%%%%%%%%%%%%%%%%%%%%%%%%%
%%%%%%%%%%%%%%%%%%%%%%%%%%%%%%%%%%%%%%%%%%%%%%%%%%%%%%%%%%%%%%%%%%%%%%%%%%%%%%%%%%%%%%%%%%%%%%
\subsection{Non-perfect spinorial fluid}
In this case, both the momentum density $q^{i}=-Qs^i$ and the anisotropic pressure $\Pi^{ij}=\Pi\left(s^is^j+\frac{1}{2}N^{ij}\right)$ can be different from zero. The expressions of $\Pi$ and $Q$ are given by eqs. \eqref{thermo_Pi} and \eqref{thermo_Q}. Following the lines of Section \ref{Section3}, we can distinguish two main cases: $\Omega\xi=0$ and $\Omega\xi\neq0$. In this paper, we focus only on the case $\Omega\xi=0$, so on LRS space-times of type I, II, and III. We are currently studying the case $\Omega\xi\neq0$, which poses some difficulties of both a conceptual and technical nature. Our future findings will be presented in a forthcoming paper. 
\\
\\
{ \bf LRSI}: $\Omega \neq0$, $\xi=0$.
\\
Condition \eqref{rif} implies $\dot{f}=0$ for every covariant scalar $f$. As in the case of a perfect fluid, from equations \eqref{dotomega} and \eqref{hatxi} we have $\Sigma=\Theta=0$. Again, eq. \eqref{rif2} implies $q_i=0$ which is consistent with $\dot\beta=\xi=0$.
The covariant equations \eqref{dotomega}, \eqref{dot_H}, \eqref{dot_phi}, \eqref{hatxi}, \eqref{hat_Theta_Sigma}, \eqref{dot_mu} and \eqref{dot_E} are automatically satisfied, as well as the constraint \eqref{dotA} and the Dirac equation involving $\dot\rho$. The evolution equation \eqref{dot_xi} reduces to the constraint \eqref{constraint_H}, giving us the expression for $H$. 

That said, by combining eq. \eqref{Raychaudhuri1+1+2} with eq. \eqref{dot_Sigma}, we get the following expression for $E$
\bea\label{eexpr} 
E=-A\phi+2\Omega^2-\frac{1}{3}\left(\mu+3p\right)-\frac{1}{2}\Pi 
\eea
in terms of $A$, $\phi$, $\Omega$, $\rho$ and $\beta$. In connection with this, a direct calculation shows that the expressions \eqref{constraint_H} and \eqref{eexpr} are consistent with the propagation equations \eqref{hat_E} and \eqref{hat_H}: by inserting \eqref{constraint_H} and \eqref{eexpr} into eqs. \eqref{hat_E} and \eqref{hat_H}, we get two automatically satisfied identities. Moreover, by replacing the expression \eqref{eexpr} into the propagation equation \eqref{hat_phi}, we obtain the equation
\be\label{phi21}
\hat{\phi}= -\frac{1}{2}\phi^{2}-A\phi+2\Omega^2-\left(\mu+p\right)-\Pi
\ee
Summing it all up, we are left with the following set of differential equations
\bea\label{eq_finali_LRSI_non_perfect}
\begin{cases}
\hat{A}=-A\phi+A^2+2\Omega^2-\frac{1}{2}\left(\mu+3p\right) \\
\hat{\Pi} + \hat{p} = - \frac{3}{2}\Pi\phi + \Pi\/A + A\left(\mu+p\right) \\
\hat{\Omega}=-\Omega\left(A+\phi\right) \\
\hat{\phi}= -\frac{1}{2}\phi^{2}-A\phi+2\Omega^2-\left(\mu+p\right)-\Pi \\
\hat{\rho}=\rho\left(2m\sin{\beta}+A-\phi\right)
\end{cases}
\eea
for the unknowns $A$, $\Omega$, $\phi$, $\rho$ and $\beta$, and where
\bea
\begin{cases}
\mu=\frac{1}{2}\rho\left(m\cos{\beta}-\Omega-\frac{1}{2}\hat{\beta}\right)\\
p=-\frac{1}{12}\rho\left(\hat{\beta}+2\Omega\right)\\
\Pi=\frac{1}{6}\rho\left(\Omega-\hat{\beta}\right)
\end{cases}
\eea
Because of the definitions \eqref{thermoquantities}, the second of the equations \eqref{eq_finali_LRSI_non_perfect} is a differential equation of the second order in the variable $\beta$. Clearly, it can be reduced to a set of first--order differential equations by introducing an additional variable $\alpha:=\hat\beta$. The system \eqref{eq_finali_LRSI_non_perfect} is then well--posed. Assigned suitable initial data for the unknowns, it admits (at least locally) a unique solution. 
\\
\\
{\bf LRSII}: $\Omega=0$ and $\xi=0$
\\
There are no constraints on the {\it dot} and {\it hat} derivatives of covariant scalar quantities. Additionally, we have no restrictions on the momentum density $q_{i}$ and the anisotropic pressure $\Pi_{ij}$. We immediately have
\bea 
H=0 
\eea
The evolution equations of \eqref{dotomega}, \eqref{dot_H}, \eqref{dot_xi}, and the propagation equations of \eqref{hat_Omega}, \eqref{hatxi} and \eqref{hat_H}, are identically satisfied. The remaining covariant equations, together with the constraint \eqref{dotA} and the Dirac equations, are given by
\begin{subequations}\label{ray_LRSII_np}
\begin{align}
&\dot{\Theta}+A\phi +\hat{A}- A^{2}+\frac{1}{3}\Theta^{2}+\frac{3}{2}\Sigma^{2}+\frac{1}{2}\left(\mu+3p\right)=0\label{ray12}\\
&\dot{\mu} +\left(\mu + p\right)\Theta -\hat{Q}-Q\phi+2AQ-\frac{3}{2}\Sigma\Pi=0\label{mupunto}\\
& \dot{Q}-\hat{\Pi}-\frac{3}{2}\Pi\phi+\Pi A-\hat{p}+A\left(\mu + p\right)+\frac{4}{3}\Theta Q-Q\Sigma=0\label{qpunto}\\
&\dot{\Sigma}-\frac{2}{3}\hat{A}+\frac{1}{3}A\phi-\frac{1}{2}\Sigma^{2}+\frac{2}{3}A^{2}+\frac{2}{3}\Theta\Sigma+E+\frac{1}{2}\Pi=0 \label{sigmapunto}\\
&\dot{\phi}+\left(\Sigma+\frac{2}{3}\Theta\right)\left(A+\frac{1}{2}\phi\right)+Q=0\label{phipunto}\\
& -\dot{E}-\Theta E-\frac{3}{2}E\Sigma+\frac{1}{2}\dot{\Pi}+\frac{2}{3}AQ=\frac{1}{3}\hat{Q}-\frac{1}{6}Q\phi+\frac{1}{2}\left(\mu+p\right)\Sigma-\frac{1}{6}\Theta\Pi+\frac{1}{4}\Pi\Sigma\label{epunto}\\
&-\frac{2}{3}\hat{\Theta}-\hat{\Sigma}-\frac{3}{2}\Sigma\phi-Q=0\label{thetacap}\\
&\hat{\phi}+\frac{1}{2}\phi^2+\left(\Sigma-\frac{1}{3}\Theta\right)\left(\Sigma+\frac{2}{3}\Theta\right)+\frac{2}{3}\mu+\frac{1}{2}\Pi-E=0\label{phicap}\\
&\hat{E}-\frac{1}{2}\hat{\Pi}+\frac{1}{3}\hat{\mu}+\frac{3}{2}\phi\left(E-\frac{1}{2}\Pi\right)+Q\left(\frac{1}{3}\Theta+\frac{1}{2}\Sigma\right)=0\label{ecap}\\
&\dot{A}= \non - 2m\dot{\beta}\cos{\beta} - \left(\Sigma+\frac{2}{3}\Theta\right)\left(A+\frac{1}{2}\phi\right) - Q - \hat{\Theta} + A\Theta + \\
& +\left(\Sigma-\frac{1}{3}\Theta\right)\left(2m\sin{\beta}+A-\phi\right) \\
&\dot{\ln{\rho}}+\Theta=0\label{rhopunto}\\
&\hat{\ln{\rho}}=\!2m\sin{\beta}+A-\phi \label{rhocap}
\end{align}
\end{subequations}
where now we have
\bea\label{Q_Pi_LRSII_np} \begin{cases}
Q&=-\frac{1}{4}\rho\dot{\beta}\\
\Pi&=-\frac{1}{6}\rho\hat{\beta}\\
\mu&=\frac{1}{2}\rho\left(m\cos{\beta}-\frac{1}{2}\hat{\beta}\right) = \frac{1}{2}\rho\/m\cos\beta + \frac{3}{2}\Pi \\
p&=-\frac{1}{12}\rho\hat{\beta}=\frac{1}{2}\Pi
\end{cases}
\eea
From the expressions of $Q$ and $\Pi$ in eq. \eqref{Q_Pi_LRSII_np}, we have the relations
\be\label{dothatbeta}
\dot{\beta}=-\frac{4}{\rho}Q \quad {\rm and} \quad \hat{\beta}=-\frac{6}{\rho}\Pi 
\ee
In view of this, we can treat $Q$ and $\Pi$ as independent variables and consider eqs. \eqref{dothatbeta} as the evolution and propagation equations for the chiral angle $\beta$. If we choose to follow this idea, we must verify the integrability conditions (see eq. \eqref{comts}) 
\bea\label{intbeta} 
\hat{\dot{\beta}} - \dot{\hat{\beta}}= + A\dot{\beta} - \Sigma\hat{\beta} + \frac{1}{3}\Theta\hat{\beta} 
\eea
for eqs. \eqref{dothatbeta}. In this regard, from the \eqref{dothatbeta} we get 
\bea\label{intbeta1}  
\hat{\dot{\beta}}=\frac{4}{\rho^2}\hat{\rho}Q-\frac{4}{\rho}\hat{Q} \quad {\rm and} \quad \dot{\hat{\beta}}=\frac{6}{\rho^2}\dot{\rho}\Pi-\frac{6}{\rho}\dot{\Pi} 
\eea
Inserting eqs. \eqref{dothatbeta} and \eqref{intbeta1}  into eq. \eqref{intbeta} and making use of the Dirac equations \eqref{rhopunto} and \eqref{rhocap}, we obtain the final equation
\bea\label{condizioni_integrabilità_dothatbeta}
\label{intbeta2} 2Q\left(-\phi+2m\sin{\beta}\right)-2\hat{Q}+4\Theta\Pi+3\dot{\Pi} + 4QA - 3\Sigma\Pi =0
\eea
Now, eq. \eqref{condizioni_integrabilità_dothatbeta} results in being equivalent to eq. \eqref{mupunto}. Indeed, taking the identities $\mu=\frac{1}{2}\rho\/m\cos{\beta}+\frac{3}{2}\Pi$ and $\mu + p =\frac{1}{2}\rho\/m\cos{\beta}+2\Pi$ into account, it is easily seen that eq. \eqref{mupunto} is identical to eq. \eqref{condizioni_integrabilità_dothatbeta} up to a multiplication factor $1/2$: 
\bea 
\non &\left(\frac{1}{2}\dot{\rho}m\cos{\beta}-\frac{1}{2}\rho\/m\dot{\beta}\sin{\beta}+\frac{3}{2}\dot{\Pi}\right)+\Theta\left(\frac{1}{2}\rho\/m\cos{\beta}+2\Pi\right)-\hat{Q}-Q\phi+2AQ-\frac{3}{2}\Sigma\Pi = \\
&\label{mupunto1}  = 2mQ\sin{\beta}+\frac{3}{2}\dot{\Pi}+2\Theta\Pi-\hat{Q}-Q\phi+2AQ-\frac{3}{2}\Sigma\Pi=0 
\eea
The integrability conditions of eqs. \eqref{dothatbeta} are then ensured by eq. \eqref{mupunto}. To conclude, making use of eqs. \eqref{Q_Pi_LRSII_np}, eqs. \eqref{ray_LRSII_np} and \eqref{dothatbeta} can be recast in the final form
\bea\label{sistema_finale_LRSII_np}
\begin{cases}
\dot{\Theta}+A\phi +\hat{A}- A^{2}+\frac{1}{3}\Theta^{2}+\frac{3}{2}\Sigma^{2}+\frac{1}{2}\left(\frac{1}{2}\rho\/m\cos\beta + 3\Pi\right)=0 \\
\frac{3}{2}\dot{\Pi} - \hat{Q} + 2mQ\sin{\beta}+2\Theta\Pi-Q\phi+2AQ-\frac{3}{2}\Sigma\Pi=0 \\
\dot{Q}-\frac{3}{2}\hat{\Pi}-\frac{3}{2}\Pi\phi + 3\Pi\/A + \frac{1}{2}A\rho\/m\cos\beta + \frac{4}{3}\Theta\/Q - Q\Sigma=0 \\
\dot{\Sigma}-\frac{2}{3}\hat{A}+\frac{1}{3}A\phi-\frac{1}{2}\Sigma^{2}+\frac{2}{3}A^{2}+\frac{2}{3}\Theta\Sigma+E+\frac{1}{2}\Pi=0 \\
\dot{\phi}+\left(\Sigma+\frac{2}{3}\Theta\right)\left(A+\frac{1}{2}\phi\right)+Q=0 \\
\dot{E} + E\Theta + \frac{3}{2}E\Sigma + \frac{1}{2}\Theta\Pi + \frac{2}{3}mQ\sin\beta - \frac{1}{2}Q\phi + 
\frac{1}{4}m\rho\cos\beta\/\Sigma + \frac{3}{4}\Sigma\Pi =0 \\
\frac{2}{3}\hat{\Theta}+\hat{\Sigma}+\frac{3}{2}\Sigma\phi+Q=0 \\
\hat{\phi}+\frac{1}{2}\phi^2+\left(\Sigma-\frac{1}{3}\Theta\right)\left(\Sigma+\frac{2}{3}\Theta\right)+\frac{1}{3}m\rho\cos\beta 
+ \frac{3}{2}\Pi-E=0 \\
\hat{E}+\frac{1}{6}m\rho\cos{\beta}\left(A-\phi+2m\sin{\beta}\right)+m\Pi\sin{\beta}+\frac{3}{2}\phi\left(E-\frac{1}{2}\Pi\right)+Q\left(\frac{1}{3}\Theta+\frac{1}{2}\Sigma\right)=0\\
\dot{A} + \hat{\Theta} + 2m\dot{\beta}\cos{\beta} + \left(\Sigma+\frac{2}{3}\Theta\right)\left(A+\frac{1}{2}\phi\right) + Q - A\Theta - \left(\Sigma-\frac{1}{3}\Theta\right)\left(2m\sin{\beta}+A-\phi\right) =0 \\
\dot{\ln{\rho}}+\Theta=0 \\
\hat{\ln{\rho}}= 2m\sin{\beta}+A-\phi \\
\dot{\beta}=-\frac{4}{\rho}Q \\ 
\hat{\beta}=-\frac{6}{\rho}\Pi 
\end{cases}
\eea
The differential system \eqref{sistema_finale_LRSII_np} consists of fourteen equations for nine unknowns. Given the consistency conditions discussed in Section \ref{Section3}, along with the integrability conditions for the Dirac equations \eqref{rhopunto} and \eqref{rhocap} and the equations \eqref{dothatbeta}, a general solution algorithm for the system \eqref{sistema_finale_LRSII_np} can be borrowed from \cite{Ellisperf} and it is as follows. Given a space-like hypersurface $\sigma$, orthogonal to $u^i$, the variables $A$, $\Theta$, $Q$, and $\Pi$ can be freely specified on $\sigma$. The seventh, eighth, ninth, twelfth, and fourteenth equations of \eqref{sistema_finale_LRSII_np} are then used to determine the spatial distribution on $\sigma$ of the variables $\Sigma$, $\phi$, $E$, $\rho$, and $\beta$, respectively. All remaining equations provide us with the expressions for the time derivatives on $\sigma$ of all nine unknowns, from which the evolution of the unknowns along the time-like congruence follows.

Another idea to solve the system \eqref{sistema_finale_LRSII_np} might be to freely choose some variables to obtain a simpler system of consistent propagation and evolution equations for the remaining unknowns. An example of such an approach is given in Section \ref{Section6}, where we present an exact integration of eqs. \eqref{sistema_finale_LRSII_np}, related to an a priori choice of some variables.
\\
\\
{\bf LRSIII}: $\Omega=0$, $\xi\neq0$.
\\
Requirement \eqref{rif} implies $\hat{f}=0$ for every covariant scalar $f$, so in particular $\hat\beta=0$. In view of eqs. \eqref{thermo_Pi} and \eqref{EMTLRS}, we have necessarily $\Pi_{ij}=0$. Moreover, due to the constraint \eqref{rif2}, the momentum density is zero too: $q_i=0$. The spinorial fluid is then forced to be perfect. We therefore return to the case that we have already discussed in the previous Subsection, with the same conclusion: there are no solutions of LRSIII type.
%%%%%%%%%%%%%%%%%%%%%%%%%%%%%%%%%%%%%%%%%%%%%%%%%%%%%%%%%%%%%%%%%%%%%%%%%%%%%
%%%%%%%%%%%%%%%%%%%%%%%%%%%%%%%%%%%%%%%%%%%%%%%%%%%%%%%%%%%%%%%%%%%%%%%%%%%%%
\section{Some exact solutions}\label{Section6}
In this section, we explore some exact solutions of the differential systems discussed in Section \ref{Section5}.
\subsection{FLRW spatially flat solution}
Let us consider the system \eqref{condizioni_1.a_LRSII}, which we rewrite below for the convenience of the reader
\bea\label{condizioni_1.a_LRSII_Sec6} 
\begin{cases}
\frac{1}{3}\Theta^2 - \mu=0 \\
\dot{\Theta}+\frac{1}{2}\Theta^2=0 \\
\dot{\ln{\rho}} + \Theta=0
\end{cases}
\eea
Let us also remember the following conditions
\be\label{condizioni_esempio1}
\beta=0, \quad A=\Omega=\Sigma=\phi=\xi=E=H=0 \quad {\rm and} \quad \hat{f}=0 \ \ \forall \ \ {\rm covariant \ \ scalar} \ \ f
\ee
which have been employed to deduce the final equations \eqref{condizioni_1.a_LRSII_Sec6}. In particular, assumptions \eqref{condizioni_esempio1} imply that the worldlines of the time-like congruence are geodesic and surface--orthogonal, the space-time is isotropic, homogeneous, and conformally flat. Such a set of requirements is certainly met by adopting an FLRW spatially flat metric
\bea\label{FLRW} 
ds^2=dt^2-a^{2}(t)\left( dx^2+dy^2+dz^2\right) 
\eea
and setting $u^i=\delta^i_t$, in such a way that the vector field $u^i$ coincides with the vector field $\frac{\partial}{\partial t}$. In connection with this, it is a straightforward matter to verify that eqs. \eqref{condizioni_1.a_LRSII_Sec6} exactly reproduces the content of the Einstein--Dirac equations, after evaluating them in the metric \eqref{FLRW} and appropriately choosing the form of the spinor field.

To see this point, choosing the Dirac representation for a set of Clifford matrices $\gamma^\nu$ ($\nu=0,\ldots,3$) and making use of the following co--tetrad field  
\be\label{tetrad_FLRW}
e^0=dt, \quad e^1=a(t)\,dx, \quad e^2=a(t)\,dy, \quad e^3=a(t)\,dz
\ee
the Einstein--Dirac equations result in being expressed as
\begin{subequations}\label{Einstein_eq_es1}
\begin{equation}\label{Einstein_eq_es1a}
3\left(\frac{\dot a}{a}\right)^2 = \frac{1}{2}m\bar\psi\psi 
\end{equation}
\begin{equation}\label{Einstein_eq_es1b}
2\frac{\ddot a}{a} + \left(\frac{\dot a}{a}\right)^2 = 0
\end{equation}
\be\label{flrwdir}
\dot{\psi}+\frac{3\dot{a}}{2a}\psi+im\gamma^{0}\psi=0 
\ee
\end{subequations}
where $\psi$ is the spinor field, $\bar\psi$ its compex conjugate and $m$ is the spinor mass. 
A solution of eqs. \eqref{Einstein_eq_es1} is given by 
\bea\label{flrwspinor} 
a(t)=a_0t^{\frac{2}{3}} \quad {\rm and} \quad \psi=\frac{C}{a(t)^{\frac{3}{2}}}
\begin{pmatrix}
e^{-imt} \\
0\\
0\\
0 
\end{pmatrix}
\eea
where $a_0$ and $C$ are suitable integration constants, with $C$ complex quantity satisfying $\frac{8}{3}{a_0}^3=m|C|^2$ (from \eqref{Einstein_eq_es1a}). The spinorial current $\bar\psi\gamma^i\psi$ and the spin pseudo--vector field $\bar\psi\gamma^i\gamma^5\psi$ associated with the Dirac field \eqref{flrwspinor} are expressed as
\be\label{current_es1}
\bar\psi\gamma^i\psi=\frac{|C|^2}{a^3}\delta^i_t \quad {\rm and} \quad \bar\psi\gamma^i\gamma^5\psi=\frac{|C|^2}{a^4}\delta^i_z
\ee
in such a way that the corresponding $4$-vector fields $u^i$ and $s^i$ coincide respectively with $\frac{\partial}{\partial t}$ and $\frac{1}{a}\,\frac{\partial}{\partial z}$. Furthermore, we also have the identity $\bar\psi\gamma^5\psi =0$, which implies $\beta=0$ as a possible choice. 

Therefore, taking the non--trivial Christoffel symbols 
\begin{equation}\label{Christoffel_es1}
\Gamma_{xt}^{\;\;\;x}= \Gamma_{yt}^{\;\;\;y}= \Gamma_{zt}^{\;\;\;z} = \frac{\dot a}{a} \quad {\rm and} \quad
\Gamma_{xx}^{\;\;\;t}= \Gamma_{yy}^{\;\;\;t}= \Gamma_{zz}^{\;\;\;t}= a{\dot a}
\end{equation}
associated with the metric \eqref{FLRW} into account and working out the covariant derivatives $\nabla_iu_j$ and $\nabla_is_j$, a direct check shows that all the requirements \eqref{condizioni_esempio1} are verified. Moreover, in view of the identity $\Theta=3\frac{\dot a}{a}$, the field equations \eqref{Einstein_eq_es1} amount to the covariant ones \eqref{condizioni_1.a_LRSII_Sec6}, having solution of the form
\be\label{soluzione_esempio1}
\Theta(t)= \frac{2}{t}, \quad \rho(t)=\frac{\rho_0}{t^2} \quad {\rm and} \quad \mu(t)=\frac{4}{3t^2}
\ee
with $\rho_0 = \frac{|C|^2}{{a_0}^3}$. The conclusion follows that the LRSII space-time, singled out by eqs. \eqref{condizioni_1.a_LRSII_Sec6} (or \eqref{condizioni_1.a_LRSII}) together with the constraints \eqref{condizioni_esempio1}, describes a FLRW spatially flat space-time, filled with a spinorial dust.
%%%%%%%%%%%%%%%%%%%%%%%%%%%%%%%%%%%%%%%%%%%%%%%%%%%%%%%%%%%%%%%%%%%%%
\subsection{Bianchi-I solution}
Let us consider a partially isotropic Bianchi-I metric of the form
\bea\label{bianchi1} 
ds^2=dt^2-a(t)^2\,dx^2-a(t)^2\,dy^2-c(t)^2\,dz^2 
\eea
A natural co-tetrad field associated with the metric \eqref{bianchi1} is given by
\begin{equation}\label{6.2}
e^0=dt, \quad e^1 = a(t)\,dx, \quad e^2 = a(t)\,dy \quad e^3 = c(t)\,dz
\end{equation}
with the tetrad field, dual of \eqref{6.2}, expressed as
\begin{equation}\label{6.3}
e_0=\frac{\partial}{\partial t}, \quad e_1 = \frac{1}{a(t)}\,\frac{\partial}{\partial x}, \quad e_2 = \frac{1}{a(t)}\,\frac{\partial}{\partial y}, \quad e_3 = \frac{1}{c(t)}\,\frac{\partial}{\partial z}
\end{equation}
The non--trivial Christoffel symbols associated with the metric \eqref{bianchi1} are
\begin{equation}\label{6.4}
\Gamma_{xt}^{\;\;\;x}= \Gamma_{yt}^{\;\;\;y}=\frac{\dot a}{a}, \quad \Gamma_{zt}^{\;\;\;z}=\frac{\dot c}{c}, \quad
\Gamma_{xx}^{\;\;\;t}= \Gamma_{yy}^{\;\;\;t}=a{\dot a}, \quad \Gamma_{zz}^{\;\;\;t}= c{\dot c}
\end{equation}
Once again adopting the Dirac representation for a set of Clifford matrices $\gamma^\mu$, the spinor covariant derivative induced by the Levi--Civita connection \eqref{6.4} is expressed as
\begin{equation}\label{6.7}
\tilde{D}_i\psi = \partial_i\psi - \tilde{\Omega}_i\psi, \qquad \tilde{D}_i\bar\psi = \partial_i\bar\psi + \bar{\psi}\tilde{\Omega}_i
\end{equation}
where the spinor connection coefficients $\tilde{\Omega}_i$ are given by
\begin{equation}\label{6.8}
\tilde{\Omega}_t=0, \quad \tilde{\Omega}_x=\tilde{\Omega}_y=\frac{1}{2}{\dot a}\gamma^1\gamma^0, \quad \tilde{\Omega}_z=\frac{1}{2}{\dot c}\gamma^3\gamma^0
\end{equation} 
Taking eqs. \eqref{6.4}, \eqref{6.7} and \eqref{6.8} into account, it is easily seen that the Einstein--Dirac equations assume the form (for more details, see \cite{VFC2011})
\begin{subequations}\label{6.10}
\begin{equation}\label{6.10a}
\left(\frac{\dot a}{a}\right)^2 + 2\frac{\dot a}{a}\frac{\dot c}{c} = \frac{1}{2}m\bar\psi\psi  
\end{equation}
\begin{equation}\label{6.10c}
\frac{\ddot a}{a} + \frac{\ddot c}{c} + \frac{\dot a}{a}\frac{\dot c}{c} = 0
\end{equation}
\begin{equation}\label{6.10d}
2\frac{\ddot a}{a} + \left(\frac{\dot a}{a}\right)^2 = 0
\end{equation}
\begin{equation}\label{6.9}
\dot\psi + \frac{\dot\tau}{2\tau}\psi + im\gamma^0\psi =0 
\end{equation}
\begin{equation}\label{6.11b}
\bar\psi\gamma^5\gamma^x\psi =0
\end{equation}
\begin{equation}\label{6.11c}
\bar\psi\gamma^5\gamma^y\psi =0
\end{equation}
\end{subequations}
where we have denoted $\tau := a^2c$ and $\gamma^i=\gamma^\mu\/e_\mu^i$. A solution of the Dirac equations \eqref{6.9} which satisfies the constraints \eqref{6.11b} and \eqref{6.11c} (coming from the non--diagonal part of the Einstein equations) is again of the form
\bea\label{Bianchi1spinor} 
\psi=\frac{C}{\sqrt{\tau}}
\begin{pmatrix}
e^{-imt} \\
0\\
0\\
0 
\end{pmatrix}
\eea
where $C$ is a complex integration constant. The scalar and vector bi--linears generated by the spinor \eqref{Bianchi1spinor} are given by
\begin{equation}
\bar\psi\psi=\frac{|C|^2}{\tau}, \quad \bar\psi\gamma^5\psi=0, \quad \bar{\psi}\gamma^i\psi=\frac{|C|^2}{\tau}\delta^i_t \quad {\rm and} \quad \bar{\psi}\gamma^i\gamma^{5}\psi=\frac{|C|^2}{c\tau}\delta^i_z
\end{equation}
so that the corresponding $1$-forms $u_i$ and $s_i$ are expressed respectively as
\begin{equation}
u_i=\delta_i^t \quad {\rm and} \quad s_i=-c(t)\delta_i^z
\end{equation}
Thus, calculating the covariant derivatives $\nabla_iu_j$ and $\nabla_is_j$ and the components of Dirac energy--momentum tensor, we obtain the identities
\begin{equation}\label{6.12}
A=\Omega=\xi=\phi=H=\beta=p=Q=\Pi=0
\end{equation}
The spatial metric $h_{ij}=g_{ij}-u_iu_j$ and the bi--spatial tensor $N_{ij}=h_{ij}+s_is_j$ are here of the form
\begin{subequations}
\begin{align}\label{6.13}
h_{ij}\,dx^i\otimes dx^j &=-a^2\,\left(dx\otimes dx + dy\otimes dy\right) -c^2\,dz\otimes dz \\
N_{ij}\,dx^i\otimes dx^j &=-a^2\,\left(dx\otimes dx + dy\otimes dy\right)
\end{align}
\end{subequations}
The only covariant kinematical quantities which are not zero are the expansion scalar $\Theta=\bar\nabla_{i}u^{i}$ and the shear tensor $\sigma_{ij}=\bar{\nabla}_{(i}u_{j)}-\frac{1}{3}\bar{\nabla}_{q}u^{q}h_{ij}$. The former is given by
\bea\label{6.14}
\Theta=2\frac{\dot{a}}{a}+\frac{\dot{c}}{c} 
\eea
whereas the non vanishing components of the latter are expressed as
\be\label{6.15}
\sigma_{xx}=\sigma_{yy} = -\frac{1}{3}a\dot{a}+\frac{1}{3}a^2\frac{\dot{c}}{c} \quad {\rm and} \quad
\sigma_{zz}=-\frac{2}{3}c\dot{c}+\frac{2}{3}c^2\frac{\dot{a}}{a}
\ee
Defining the shear scalar
\be\label{6.16}
\Sigma=\sigma_{ij}s^{i}s^{j}=\sigma_{ij}N^{ij}=-\frac{2}{3}\frac{\dot{c}}{c}+\frac{2}{3}\frac{\dot{a}}{a}
\ee
the components of the shear tensor read as
\be\label{6.17}
\sigma_{ij}=\Sigma\left(s_is_j+\frac{1}{2}N_{ij}\right)
\ee
in accordance with the requirements of the LRS geometry. Now, the covariant scalar quantities $\Theta$ and $\Sigma$ have to satisfy the equations \eqref{sistema_finale_phi=0_LRSII}, namely
\begin{subequations}\label{6.18}
\begin{align}
\label{6.18a} &\dot{\Theta}+\frac{1}{3}\Theta^2+\frac{3}{2}\Sigma^2+\frac{1}{2}\mu=0 \\
\label{6.18b} &\dot{\Sigma}-\frac{1}{2}\Sigma^{2}+\frac{2}{3}\Theta\Sigma+\left(\Sigma-\frac{1}{3}\Theta\right)\left(\Sigma+\frac{2}{3}\Theta\right)+\frac{2}{3}\mu=0 \\
\label{6.18c} &\dot{\ln\rho} + \Theta=0
\end{align}
\end{subequations}
where here $\rho=\bar\psi\psi$, since $\Omega=\beta=0$.
Inserting the content of eqs. \eqref{6.14} and \eqref{6.16} into eqs. \eqref{6.18}, the latter assume the form
\begin{subequations}\label{6.19}
\begin{align}
\label{6.19a} &2\frac{\ddot{a}}{a}+\frac{\ddot{c}}{c}=-\frac{1}{4}m\bar{\psi}\psi \\
\label{6.19b} &\frac{\ddot{a}}{a}-\frac{\ddot{c}}{c}-3\frac{\dot{a}}{a}\frac{\dot{c}}{c}=-\frac{1}{2}m\bar{\psi}\psi \\
\label{6.19c} &\dot{\ln\rho}=-\dot{\ln\tau}
\end{align}
\end{subequations}
It is evident that the spinor field \eqref{Bianchi1spinor} verifies eq. \eqref{6.19c}. Moreover, a direct check shows that eqs. \eqref{6.19a} and \eqref{6.19b} are identical to suitable linear combinations of the Einstein equations \eqref{6.10}. In detail, we have: $\eqref{6.19a}=2\eqref{6.10c} + \eqref{6.10d} -\eqref{6.10a}$ and $\eqref{6.19b} = \eqref{6.10d} - \eqref{6.10c} - \eqref{6.10a}$. As a last remark, we note that eq. \eqref{efi0} is implied by eqs. \eqref{6.10} too. Indeed, making use of eq. \eqref{6.18b}, eq. \eqref{efi0} can be rewritten in the form
\bea\label{efi0bis}
E= -\dot\Sigma + \frac{1}{2}\Sigma^2 - \frac{2}{3}\Theta\Sigma
\eea
Therefore, on one side, by definition, we have the identity
\be\label{6.20}
E=C_{ijhk}u^is^ju^hs^k = \frac{1}{3}\left(\frac{\ddot c}{c} - \frac{\ddot a}{a} - \frac{\dot a}{a}\frac{\dot c}{c} + \left(\frac{\dot a}{a}\right)^2\right)
\ee
On the other side, by a direct calculation, we get
\be\label{6.21}
-\dot\Sigma + \frac{1}{2}\Sigma^2 - \frac{2}{3}\Theta\Sigma = \frac{2}{3}\left(\frac{\ddot c}{c} - \frac{\ddot a}{a}\right)
\ee
Inserting eqs. \eqref{6.20} and \eqref{6.21} into eq. \eqref{efi0bis}, we obtain the equation
\be\label{6.22}
\frac{\ddot a}{a} - \frac{\ddot c}{c} - \frac{\dot a}{a}\frac{\dot c}{c} + \left(\frac{\dot a}{a}\right)^2 =0
\ee
clearly identical to the linear combination $\eqref{6.10d} - \eqref{6.10c}$.

We conclude that the LRSII space-time, determined by the covariant equations \eqref{6.18} (or \eqref{sistema_finale_phi=0_LRSII}) and by the constraints \eqref{6.12} and \eqref{efi0bis} (or \eqref{efi0}), represents a partially isotropic Bianchi-I space-time \eqref{bianchi1}, filled with a spinorial dust.
%%%%%%%%%%%%%%%%%%%%%%%%%%%%%%%%%%%%%%%%%%%%%%%%%%%%%%%%%
\subsection{Minkowski solutions}
Let us discuss paticular solutions of the system \eqref{sistema_finale_LRSII_np} under some simplifying hypotheses. In particular, we assume $A=0$ and $\Sigma=\frac{1}{3}\Theta$. Under such conditions, the vector fields $u^i$ and $s^i$ commute (see eq. \eqref{comts}). Thus, according to the Frobenius theorem, the distribution generated by $u^i$ and $s^i$ is integrable, its integral surfaces give rise to a foliation of space-time, the parameters of the time-like and space-like congruences can be assumed as local coordinates on the leaves, and possibly completed to local coordinates over the entire space-time.

Under the assumptions $A=0$ and $\Sigma=\frac{1}{3}\Theta$, the covariant equations \eqref{sistema_finale_LRSII_np} become
\begin{subequations}\label{covEq.6}
\begin{align}
&4\dot{\Theta}+6\Pi+2\Theta^2+m\rho\cos{\beta}=0\label{thetapunto1.6}\\
&3(\dot{\Pi}+\Pi\Theta)+4mQ\sin{\beta}=2(\hat{Q}+Q\phi)\label{pipunto.6}\\
&\dot{Q}+Q\Theta=\frac{3}{2}(\hat{\Pi}+\Pi\phi)\label{qpunto.6}\\
&2\dot{\Theta}+6E+3\Pi+\Theta^2=0\label{thetapunto2.6}\\
&\dot{\phi}+Q+\frac{1}{2}\Theta\phi=0\label{phipunto.6}\\
&12\dot{E}+18E\Theta+9\Pi\Theta+m\Theta\rho\cos{\beta}+8mQ\sin{\beta}=6Q\phi\label{epunto.6}\\
&\hat{\Theta}+Q+\frac{1}{2}\Theta\phi=0\label{thetacap.6}\\
&6\hat{\phi}+9\Pi+3\phi^2-6E+2m\rho\cos{\beta}=0\label{phicap.6}\\
&3(4\hat{E}+2Q\Theta+6E\phi -3\Pi\phi+4m\Pi\sin{\beta})+2m\rho\cos{\beta}(-\phi+2m\sin{\beta})=0\label{ecap.6}\\
&\hat{\Theta}+Q+\frac{1}{2}\Theta\phi+2m\dot{\beta}\cos{\beta}=0\label{thetacap2.6}\\
&\dot{\rho}+\rho\Theta=0\label{rhopunto.6}\\
&\hat{\rho}=2m\rho\sin{\beta}-\rho\phi\label{rhocap.6}\\
&\dot{\beta}=-\frac{4}{\rho}Q\label{betapunto.6}\\
&\hat{\beta}=-\frac{6}{\rho}\Pi\label{betacap.6}
\end{align}
\end{subequations}
The compatibility between eq. \eqref{thetacap.6} and eq. \eqref{thetacap2.6} implies
\bea\label{condizione_beta_es4} 
\dot{\beta}\cos{\beta}=0 \quad \iff \quad \dot{\beta}=0 \cup \beta=\frac{\pi}{2}+k\pi 
\eea
Condition $\dot{\beta}=0$ (with $\beta \not=\frac{\pi}{2}+k\pi$) is not allowed (details are omitted for brevity), so we focus on the case $\beta=\frac{\pi}{2}+k\pi$. From eqs. \eqref{betapunto.6} and \eqref{betacap.6}, we get $Q=0$ and $\Pi=0$. From eqs. \eqref{thetapunto1.6}  and \eqref{thetapunto2.6}, we obtain $E=0$. Moreover, we have $\mu=0$ and $p=0$ as well. Therefore, the energy--momentum tensor of the Dirac field is zero, although the Dirac field itself may be non-zero. Both the Weyl and Ricci tensors are zero, and then the space-time is flat. After denoting by $\eta$ and $\chi$ the curvilinear abscissas of the time-like and space-like congruences respectively, the system of equations \eqref{covEq.6} reduces to
\begin{subequations}\label{coveq2.6}
\begin{align}
&\frac{\partial \Theta}{\partial \eta} + \frac{1}{2} \Theta^2 = 0\label{1.6} \\
 &\frac{\partial \phi}{\partial \eta} + \frac{1}{2} \Theta \phi = 0 \label{2.6}\\
 &\frac{\partial \Theta}{\partial \chi} + \frac{1}{2} \Theta \phi = 0\label{3.6} \\
&\frac{\partial \phi}{\partial \chi} + \frac{1}{2} \phi^2 = 0\label{4.6} \\
 &\frac{\partial \rho}{\partial \eta} + \rho \Theta = 0 \label{5.6}\\
 &\frac{\partial \rho}{\partial \chi}= \pm 2m\rho - \rho\phi \label{6.6}
\end{align}
\end{subequations}
for the three unknown functions $\Theta(\eta,\chi)$, $\phi(\eta,\chi)$, and $\rho(\eta,\chi)$. In eq. \eqref{6.6} the sign $+$ is related to the choice $\beta=\frac{\pi}{2}+2k\pi$, whereas the sign $"-"$ sign comes from $\beta=\frac{3}{2}\pi+2k\pi$. By first solving the equations for $\Theta$ and $\phi$ and then solving the equations for $\rho$, we get the final solutions
\begin{equation}\label{coveq3.6}
\begin{cases}
&\Theta(\eta, \chi) = 3\Sigma(\eta, \chi) = \frac{2}{\eta - C\chi+D} \\
\\
&\phi(\eta, \chi) = -\frac{2C}{\eta - C\chi+D} \\
\\
&\rho(\eta, \chi) = \frac{Ke^{\pm 2m\chi}}{(\eta- C\chi+D)^2}
\end{cases}
\end{equation}
where $C\not=0$, $D$ and $K>0$ are integration constants. Solutions \eqref{coveq3.6} have singularities on the hypersurface $\eta - C\chi+D=0$ and blow up for $\chi \to \pm \infty$. In the rest-frame and spin eingenstate ($\boldsymbol{L}\!=\!\mathbb{I}$), and by using the Chiral representation for the Clifford matrices, the explicit form of the spinor field is obtained from eq. \eqref{spinor} and is given by
\bea\label{Minkowski_spinor2} 
\psi=-\frac{\sqrt{K}e^{\pm m\chi}}{2|\eta -C\chi +D|}
\begin{pmatrix}
1-i \\
0\\
1+i\\
0 
\end{pmatrix}
\eea
We note that in the static case and setting $\Theta=0$, the system \eqref{coveq2.6} reduces to the system \eqref{condizioni_1.b_LRSII} with solution given by
\begin{equation}\label{solution_condizioni_1.b_LRSII}
\begin{cases}
&\phi(\chi) = \frac{2}{\chi+D} \\
\\
&\rho(\chi) = \frac{Ke^{\pm 2m\chi}}{(\chi+D)^2}
\end{cases}
\end{equation} 
Solutions \eqref{coveq3.6} and \eqref{solution_condizioni_1.b_LRSII} describe very particular spinor fields with a vanishing energy-momentum tensor, filling a flat space-time. Although critical from a physical point of view, such solutions are allowed by mathematics. As already mentioned, in (cylindrical) coordinates and in the static case, solutions of this kind have already been found in \cite{Critical_solutions}.
%%%%%%%%%%%%%%%%%%%%%%%%%%%%%%%%%%%%%%%%%%%%%%%%%%%%%%%%%%%%%%%%%
%%%%%%%%%%%%%%%%%%%%%%%%%%%%%%%%%%%%%%%%%%%%%%%%%%%%%%%%%%%%%%%%%
\section{Conclusion}\label{section7}
By combining the polar decomposition with the covariant approach, we developed a covariant formulation for a self-gravitating Dirac field in LRS space-times of types I, II, and III. In such a formulation, the Dirac field was described entirely in hydrodynamic form as an effective spinorial fluid, without resorting to the tetrad formalism or even to the use of Dirac matrices and their particular representations. All covariant equations were preliminarily reformulated in the signature $(+---)$, and the $(1+1+2)$ decomposition of the energy--momentum tensor of the spinor field, as well as of the Dirac equations, was carried out. By identifying the velocity and spin of the spinor field as the generators of the time-like and space-like congruences required for the $(1+1+2)$ covariant splitting, we were able to examine the Dirac field in backreaction with LRS geometries. 

Within this framework, a first finding was that if the spinor fluid is of the perfect type, only LRS space-times of types I, II, or III result to be compatible with the Dirac field. Conversely, if the spinor fluid is non-perfect, more general space-times -- beyond LRS types I, II, and III -- may become admissible. A more detailed analysis then showed that LRSIII space-times are automatically ruled out in both the perfect and non-perfect spinorial fluid cases. LRSI space-times are possible in the case of a non-perfect spinorial fluid. Instead, in the case of a perfect spinorial fluid in an LRSI space-time, the number of the resulting equations exceeds that of the unknowns, generating a constraint algorithm for which we are not able to say whether it stabilizes or not. As for LRSII space-times, we proved the existence of LRSII solutions which, in the case of perfect spinorial fluid, must necessarily be of the form of dust. 

Although it may seem natural at first to identify the time-like and space-like congruences, respectively, with the integral curves of the velocity $u^i$ and spin $s^i$ of the Dirac field, this choice can be restrictive and may be the source of many of the obstructions we encountered. In this regard, a wider choice -- still within the framework of LRS space-times -- could be to select the tangent vectors to the two congruences as coplanar with the vector fields $u^i$ and $s^i$, but not coincident with them. Moreover, space-times not belonging to LRS types I, II, and III remain to be investigated. We will devote future papers to these further lines of research, as well as to the study of general solutions of the covariant equations we obtained.

Another possible avenue for extending the present research is to go beyond LRS space-times, while still working within the 
$(1+1+2)$ decomposition framework. This line of investigation will also be pursued in future work.
%%%%%%%%%%%%%%%%%%%%%%%%%%%%%%%%%%%%%%%%%%%%%%%%%%%%%%%%%%%%%%%%%
%%%%%%%%%%%%%%%%%%%%%%%%%%%%%%%%%%%%%%%%%%%%%%%%%%%%%%%%%%%%%%%%%
\\
\\
\\
{\bf Data availability}. There is no data available in a repository.
\\
\\
{\bf Conflict of interest}. The author declares no conflict of interest.
%%%%%%%%%%%%%%%%%%%%%%%%%%%%%%%%%%%%%%%%%%%%%%%%%%%%%%%%%%%%%%%%%
%%%%%%%%%%%%%%%%%%%%%%%%%%%%%%%%%%%%%%%%%%%%%%%%%%%%%%%%%%%%%%%%%

\end{document}